\documentclass[%        	Class options:
aps,%                   	American Physical Society
prd,%                   	Physical Review D
superscriptaddress,%    	Authors' addresses linked with superscripts
nofootinbib,%           	Does not treat footnotes as references
floatfix]%              	Fixes float errors
{revtex4}%              	REVTEX 4 Package used
\usepackage{graphicx,%  	Default Latex 2eps package for embedding figures
longtable,%             	Useful for long table
color,%					Colours
bm}%						Bold math symbols
\begin{document}
%======================================================================================
\title{Global constraints on absolute neutrino masses and their ordering}
%--------------------------------------------------------------------------------------
%
\author{        	Francesco~Capozzi}
\affiliation{   	Department of Physics, Ohio State University,  Columbus, OH 43210, USA}% 191 West Woodruff Avenue,
\author{			Eleonora Di Valentino}
\affiliation{	Institut d'Astrophysique de Paris (UMR7095: CNRS \& UPMC-Sorbonne Universit{\'e}s), F-75014 Paris, France}
\affiliation{	Sorbonne Universit{\'e}s, Institut Lagrange de Paris (ILP), F-75014 Paris, France}
\author{        	Eligio~Lisi}
\affiliation{   	Istituto Nazionale di Fisica Nucleare, Sezione di Bari, %\\
               	Via Orabona 4, 70126 Bari, Italy}
\author{        	Antonio~Marrone}
\affiliation{   	Dipartimento Interateneo di Fisica ``Michelangelo Merlin,'' %\\
               	Via Amendola 173, 70126 Bari, Italy}%
\affiliation{   	Istituto Nazionale di Fisica Nucleare, Sezione di Bari, %\\
               	Via Orabona 4, 70126 Bari, Italy}
\author{			Alessandro~Melchiorri}
\affiliation{	Dipartimento di Fisica, Universit{\`a} di Roma ``La Sapienza,'' P.le Aldo Moro 2, 00185 Rome, Italy}
\affiliation{   	Istituto Nazionale di Fisica Nucleare, Sezione di Roma~I, %\\
               	P.le Aldo Moro 2, 00185 Rome, Italy}
\author{        	Antonio~Palazzo}
\affiliation{   	Dipartimento Interateneo di Fisica ``Michelangelo Merlin,'' %\\
               	Via Amendola 173, 70126 Bari, Italy}%
\affiliation{   	Istituto Nazionale di Fisica Nucleare, Sezione di Bari, %\\
               	Via Orabona 4, 70126 Bari, Italy}
\begin{abstract}%.......................................................................
\medskip
\medskip
Within the standard three-neutrino framework, the absolute neutrino masses and their ordering (either normal, NO, or inverted, IO) 
are currently unknown. However, the combination of current data coming from oscillation experiments, neutrinoless double beta 
($0\nu\beta\beta$) decay searches, and cosmological surveys, can provide interesting constraints for such unknowns in the 
sub-eV mass range, down to $O(10^{-1})$~eV in some cases.
We discuss current limits on absolute neutrino mass observables by performing a global data analysis, that includes the
latest results from oscillation experiments, $0\nu\beta\beta$ decay bounds from the KamLAND-Zen experiment, and 
constraints from representative combinations of Planck measurements and other cosmological data sets. 
In general, NO appears to be somewhat favored with 
respect to IO at the level of $\sim\!2\sigma$, mainly by neutrino oscillation data (especially atmospheric), corroborated by cosmological data
in some cases. Detailed constraints are obtained via the $\chi^2$ method,
by expanding the parameter space either around separate minima in NO and IO, or around the absolute minimum in any ordering.       
Implications for upcoming oscillation and non-oscillation neutrino experiments, including $\beta$-decay searches, 
are also discussed. 
\end{abstract}%.........................................................................
\medskip
\medskip
\maketitle

%%%%%%%%%%%%%%%%%%%%%%%%%%%%%%%%%%%%%%%%%%%%%%%
\section{Introduction}
%%%%%%%%%%%%%%%%%%%%%%%%%%%%%%%%%%%%%%%%%%%%%%%

Neutrino oscillation experiments have established that the three known flavor states $\nu_\alpha$ $(\alpha=e,\,\mu,\,\tau)$
are linear combinations of three massive states $\nu_i$ $(i=1,\,2,\,3)$ with different masses $m_i$,  
via a mixing matrix $U_{\alpha i}$ characterized by three nonzero angles $\theta_{ij}$ \cite{PDG2016}. 
Flavor oscillation frequencies in vacuum are governed by
the squared mass differences $\Delta m^2_{ij}$, that can be expressed in terms of two independent parameters, 
conventionally chosen herein as \cite{Glob06}: 
%..............................................
\begin{eqnarray}
\label{delta1}
\delta m^2 &=& m^2_2-m^2_1>0\ ,\\
\label{delta2}
\Delta m^2 &=& m^2_3-(m^2_2+m^2_1)/2\ ,
\end{eqnarray}
%..............................................
where $\Delta m^2$ can be either positive or negative according to the so-called normal ordering (NO) or inverted ordering 
(IO) for the neutrino mass
spectrum, respectively. 
Probing the mass ordering is an important goal of future experimental $\nu$ oscillation searches 
(see, e.g., \cite{Patt15,Qian15}), with relevant implications on theoretical models 
for neutrino mass and mixing (see, e.g., \cite{Moha06,Feru15,King15}).

At present, the four parameters $\delta m^2$, $|\Delta m^2|$, $\sin^2\theta_{12}$, and $\sin^2\theta_{13}$ have been measured at 
the few~\% level, while
$\sin^2\theta_{23}$ (still affected by an octant ambiguity \cite{Octa96}) 
is less accurately known, at the level of $\sim\!10\%$ \cite{PDG2016}. 
Interestingly, the combination of various oscillation data starts to show some sensitivity to the remaining unknowns, 
namely, the sign of $\Delta m^2$ and a possible CP-violating phase $\delta$, mainly through subleading
$\nu_\mu\to\nu_e$ oscillation effects in atmospheric and accelerator neutrino experiments, constrained by reactor data 
\cite{Capo16,Marr16}; 
see also \cite{Conc17,Vall14} for independent analyses of oscillation data and for discussions of the associated parameters. 

The absolute $\nu$ masses are also unknown. Lower bounds are set by oscillation data by zeroing
the lightest $m_i$,  
%...................
\begin{equation}
\label{lightest}
(m_1,\,m_2,\,m_3) \geq \left\{
\begin{array}{lr}
\left(0,\,\sqrt{\delta m^2},\,\sqrt{|\Delta m^2|+\delta m^2/2}\right) & \rm{(NO)}\ ,\\[8pt]
\left(\sqrt{|\Delta m^2|-\delta m^2/2},\,\sqrt{|\Delta m^2|+\delta m^2/2},\,0\right)  & \rm{(IO)\ .}\\
\end{array}\right.
\end{equation}
%....................
while upper bounds (and prospective measurements) can only be set by
nonoscillation neutrino experiments. 
In particular, three main observables can probe the absolute mass spectrum: $(i)$ the effective neutrino mass 
$m_\beta$ in $\beta$ decay; $(ii)$ the effective mass $m_{\beta\beta}$ in neutrinoless double beta ($0\nu\beta\beta$) 
decay, if neutrinos are Majorana fermions; and $(iii)$ the total neutrino mass $\Sigma$ in cosmology; see, e.g., the reviews in
\cite{Bile02,Otte08,Viss16,Hann10}.

These observables probe the neutrino mass spectrum in different and complementary ways \cite{PDG2016,Glob06}.  
The $\beta$ decay spectrum is sensitive to an (unresolved) combination of squared masses, 
weighted by the corresponding $\nu_e$ admixture,
%.................
\begin{equation}
\label{mb}
m_{\beta}=\sqrt{\textstyle{\sum_i |U_{ei}|^2\,m^2_i}}\ ,
\end{equation}
%.....................
while the $0\nu\beta\beta$ decay rate depends linearly on the $m_i$'s via unknown Majorana phases $\phi_i$ 
(with $\phi_1=0$ by convention),    
%
%.................
\begin{equation}
\label{mbb}
m_{\beta\beta}={\textstyle \left|\sum_i |U_{ei}|^2\,m_i\, e^{\imath\phi_i} \right|}\ ,
\end{equation}
%.....................
and cosmology essentially probes the (flavor-blind) total gravitational charge,  
%.................
\begin{equation}
\label{Sig}
\Sigma = m_1+m_2+m_3\ .
\end{equation}
%.....................

Currently, the most constraining bounds on $m_{\beta\beta}$ can be as low as $O(0.1)$~eV in the KamLAND-Zen 
experiment at $\sim 2\sigma$, 
by assuming favorable nuclear matrix elements \cite{KamZ16}. 
Upper bounds on $\Sigma$, dominated by Planck data, can also reach the level of $O(0.1)$~eV, by assuming the 
standard cosmological model \cite{Lesg14,planckparams2015}. 
Such limits are getting close to the mass scale $\sqrt{|\Delta m^2|} \simeq 0.05$~eV appearing in Eq.~(\ref{lightest})
where some sensitivity of cosmological data to mass ordering may be emerging \cite{Viss16}.
Bounds on $m_\beta$, although free from specific assumptions, are an order of magnitude weaker at present 
\cite{PDG2016,Otte08}. 
In all cases, significant improvements---and possibly a positive detection---may be expected in the next decade of 
experimental searches \cite{NuMass}.
In this context, we find it worthwhile to perform and discuss an updated global analysis of both oscillation and 
nonoscillation data, building upon previous work on the subject \cite{Melc04,Melc06,Melc08}. 

In particular, we shall highlight some
interesting features emerging from the analysis of recent data (circa 2017), namely: $(i)$ 
an increasing sensitivity to the mass ordering, with NO generally favored over IO at the $\sim2\sigma$ level; 
$(ii)$ differences in the allowed parameter space arising when such NO-IO offset is (not) taken into account;  
$(iii)$ synergies between bounds on $m_{\beta\beta}$ and $\Sigma$ of comparable strength, especially for best-fit 
values of $\Sigma$ far  
from the extrema in Eq.~(\ref{lightest}). The discussion of such features also allows to gauge the impact of 
prospective (non)oscillation
bounds---or signals---on our knowledge of the absolute neutrino mass spectrum and its associated observables.

Our work is structured as follows. In Sec.~II~A, B, and C, we report and discuss detailed constraints coming from separate
analyses of oscillations, $m_{\beta\beta}$, and $\Sigma$, respectively.  In Sec.~III we perform a combined analysis in 
the $(m_{\beta\beta},\,\Sigma)$ parameter space for representative cosmological data sets.
In Sec.~IV we discuss the implications of such results for upcoming or prospective experiments sensitive to $m_\beta$.
A brief summary is presented in Sec.~V.

\section{Data sets, statistical analysis, and parameter bounds}     

In this paper, all the bounds on the mass-mixing parameters coming from various (separate or combined) data sets
are expressed in terms of $\Delta\chi^2$ differences with respect to a minimum $\chi^2$ value. In particular, the differences 
\begin{equation}
\Delta\chi^2=n^2
\end{equation}
are used to derive $n\sigma$ allowed regions. Projections of such regions onto a single parameter provide 
the $\pm n\sigma$ range(s) for that parameter \cite{Stat16}. In all figures, it is understood that 
the undisplayed parameters are projected away
(i.e., marginalized).  Following the general statistical arguments in \cite{Blen13}, 
we use the $\Delta\chi^2$ metric also to assess the relative likelihood of the two mass-ordering hypotheses,
%---------
\begin{equation}
\label{Delta}
\Delta\chi^2_\mathrm{IO-NO}=\chi^2_{{\min},\mathrm{IO}}- \chi^2_{{\min},\mathrm{NO}}\ .
\end{equation}
%---------------

From a hystorical viewpoint, an interesting parallel to the metric in Eq.~(\ref{Delta}) can be found in 
the development of solar neutrino data analyses. 
In the early literature, different and seemingly disconnected oscillation solutions to the solar neutrino problem 
(e.g., the so-called matter and vacuum solutions) were often analyzed via
separate fits around the corresponding $\chi^2$ minima in the $(\delta m^2,\,\sin^22\theta_{12})$ parameters (see, e.g., 
\cite{Bahc98,Petc98}),
and possibly compared to each other by tests of hypotheses. However, when diverse solutions were explicitly 
connected in the $(\delta m^2,\,\sin^2\theta_{12})$ or $(\delta m^2,\,\tan^2\theta_{12})$ variables 
\cite{Mont96,Mura00,Frie00,Lisi00},
it became customary to expand the fit around the absolute $\chi^2$
minimum and to compare different solutions  by a $\Delta\chi^2$ parameter test \cite{Pena01,Gett02,PDG2016}---until a single one
was eventually found by solar and long-baseline reactor experiments \cite{Bahc04}. 

The comparison of the two mass orderings seem to follow an analogous path.
On the one hand, one may take NO and IO as two alternative options, 
involving separate fits and tests of hypotheses. On the other hand, 
one may try to connect them through a continuous variable,
involving a parameter estimation test. Such a variable could be either physical
(e.g., $\Delta m^2$, ranging from negative to positive real values) or unphysical
(e.g., a fudge parameter $p\in [0,\,1]$ linking any two competing hypotheses \cite{Lyons}). 
Explicit parametric connections 
have been worked out for medium-baseline reactor neutrino oscillations, in terms of the mixing 
variable $\sin^2\theta_{12}$ (swapping octants between NO and IO for $\Delta m^2>0$ in vacuum 
\cite{Bile17}) 
and of an empirical variable $\alpha$ (ranging in $[-1,\,+1]$ from IO to NO \cite{Juno14}). 
The above considerations further support our adoption of Eq.~(\ref{Delta}) as a reasonable metric for the IO--NO 
discrimination \cite{Blen13}, akin to a one-parameter estimation test. 
For a discussion of further statistical issues and possible alternative approaches, 
see also \cite{Patt15,Qian15,Ciuffoli,Blennow,Stanco}  and refs.\ therein.

With present data, the current statistical sensitivity associated to $\Delta\chi^2_\mathrm{IO-NO}$
tests appears to be limited to $\sim\!2\sigma$ (see Sec.~III).
Therefore, we shall conservatively report $\Delta\chi^2$ bounds on mass-mixing parameters both by  
{\em separately\/} minimizing the $\chi^2$ in NO and IO (discarding the relative $\Delta\chi^2_\mathrm{IO-NO}$ difference),
and by further minimizing the $\chi^2$ over {\em any\/} ordering (including the $\Delta\chi^2_\mathrm{IO-NO}$ information),
with a discussion of the relative differences in the results.
Such a format has been adopted in presenting the
oscillation parameter ranges in \cite{Conc17,NuFit}, and is extended herein to nonoscillation parameters.

\subsection{Neutrino oscillations}

An analysis of neutrino oscillation data has been previously presented in \cite{Capo16}, to which 
we refer the reader for a discussion of the adopted methodology and earlier literature. 
A partial update of \cite{Capo16}, 
including novel accelerator data shown in mid-2016, was reported in \cite{Marr16}.  
The more complete update presented herein (circa 2017) includes, with respect to \cite{Capo16}: 
$(i)$ the latest results from the long-baseline accelerator experiments  T2K \cite{T2K2017} and NOvA \cite{NOvA2016,NOvA2017}; 
$(ii)$ the latest far/near spectral ratio from the reactor neutrino experiment Daya Bay \cite{DB2017};  
$(iii)$ the most recent atmospheric neutrino data from the Super-Kamiokande (SK) phase IV \cite{SKatm1,SKatm2}. 
The results of our oscillation data analysis are reported graphically in Fig.~1 and 
numerically in Table~I.

Figure~1 shows the  $\chi^2$ curves in terms of the six
oscillation parameters $(\delta m^2,\,\Delta m^2,\,\sin^2\theta_{12},\,\sin^2\theta_{13},\,\sin^2\theta_{23},\,\delta)$, for both
NO (blue) and IO (red). We find an overall preference for NO, quantified by the $\chi^2$ difference
%--------------------------------
\begin{equation}
\label{preferNO}
\Delta\chi^2_\mathrm{IO-NO} = 3.6 \ \mathrm{(all\ oscill.~data)}\ ,
\end{equation}
%--------------------------------
that is explicitly shown as an offset of the IO curves. The offset is of some relevance 
in the analysis of absolute mass observables, as shown later.

%%%%%%%%%%%%%%%%%%%%%%%%%%%%%%%%%%%%%%%%%%%%%%%%%%%%%%%%%%%%%%%%%%%%%%%%%%%%%%%%%%%%%%%%%%
\begin{figure}[t]
\begin{minipage}[c]{0.96\textwidth}
\includegraphics[width=0.53\textwidth]{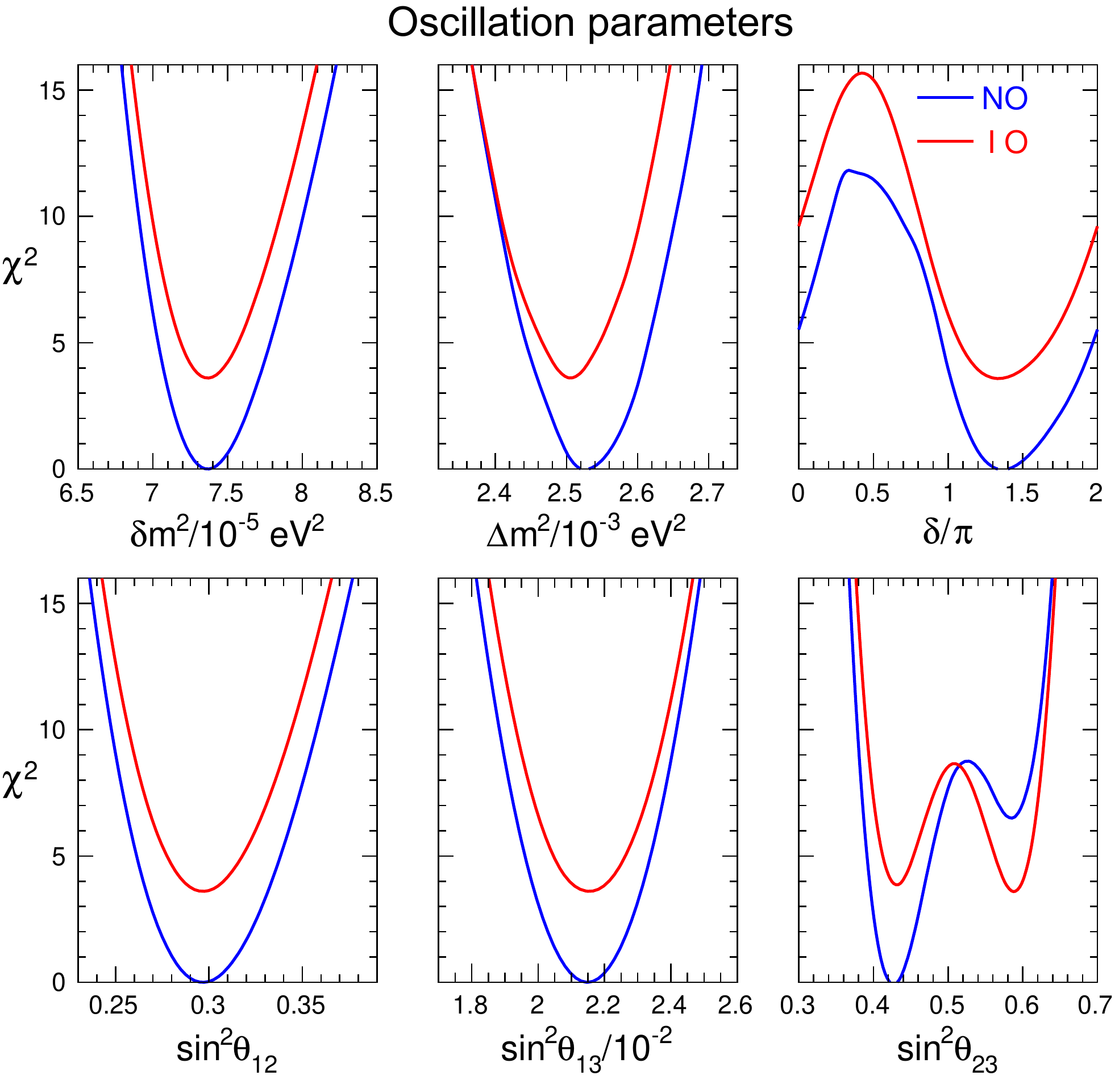}
\caption{\label{fig01}\footnotesize Global $3\nu$ oscillation analysis. Projections of the $\chi^2$ function onto the 
parameters $\delta m^2$, $|\Delta m^2|$, $\sin^2\theta_{ij}$, and $\delta$, for NO (blue) and IO (red). In each panel, 
all the undisplayed parameters are marginalized, and the offset 
$\Delta \chi^2_{\mathrm{IO}-\mathrm{NO}}= 3.6$ 
is included.} \end{minipage}
\end{figure}
%%%%%%%%%%%%%%%%%%%%%%%%%%%%%%%%%%%%%%%%%%%%%%%%%%%%%%%%%%%%%%%%%%%%%%%%%%%%%%%%%%%%%%%%%%

\newpage
Table~I reports best-fit values and parameter ranges for separate $\chi^2$ minimization 
in each separate ordering (NO and IO) and in any ordering; the latter case takes into account the above  
 $\Delta\chi^2_\mathrm{IO-NO}$ value.
The known parameters
$(\delta m^2,\,|\Delta m^2|,\,\sin^2\theta_{12},\,\sin^2\theta_{13})$, which affect the
absolute mass observables in Eqs.~(\ref{mb})--(\ref{Sig}), are determined with a fractional $1\sigma$ accuracy 
(defined as 1/6 of the $\pm3\sigma$ range) 
of $(2.3,\,1.6,\,5.8,\,4.0)$ percent, respectively. For such parameters, it turns out that minimization in any ordering
reproduces the same allowed ranges as for NO. 
Given the $\delta m^2$ and $\Delta m^2$ estimates in Table~I, Eq.~(\ref{lightest}) becomes
%...................
\begin{equation}
\label{lightest1}
(m_1,\,m_2,\,m_3) \gtrsim \left\{
\begin{array}{lc}
(0,\,0.86,\,5.06)\times 10^{-2}\ \mathrm{eV} & \rm{(NO)}\ ,\\
(4.97,\,5.04,\,0)\times 10^{-2}\ \mathrm{eV} & \rm{(IO)}\ .\\
\end{array}\right.
\end{equation}
%....................
%\vspace*{-1mm}

%===========================================================================
\begin{table}[t]
%\centering
%\captionsetup{width=.96\textwidth}
\caption{\label{Synopsis} \footnotesize Results of the global $3\nu$ oscillation analysis, in terms of best-fit values 
for the mass-mixing parameters and 
associated $n\sigma$ ranges ($n=1,\,2,\,3$), defined by $\chi^2 - \chi^2_{\min}=n^2$ with respect to the separate
minima in each mass ordering (NO, IO) and to the absolute minimum in any ordering. (Note that the fit to the 
$\delta m^2$ and $\sin^2\theta_{12}$ parameters is basically insensitive to the mass ordering.)  We recall that 
$\Delta m^2$ is defined herein
as $m^2_3-{(m^2_1+m^2_2})/2$, and that $\delta$ is 
taken in the (cyclic) interval $\delta/\pi\in [0,\,2]$. }
%\vspace*{0mm}
\centering
%\resizebox{.8\textwidth}{!}{
\begin{ruledtabular}
\begin{tabular}{lccccc}
%\hline\hline
Parameter & Ordering & Best fit & $1\sigma$ range & $2\sigma$ range & $3\sigma$ range \\
\hline%---------------------------------------------------------------------
$\delta m^2/10^{-5}~\mathrm{eV}^2 $ & NO, IO, Any & 7.37 & 7.21 -- 7.54 & 7.07 -- 7.73 & 6.93 -- 7.96 \\
\hline%---------------------------------------------------------------------
$\sin^2 \theta_{12}/10^{-1}$ & NO, IO, Any & 2.97 & 2.81 -- 3.14 & 2.65 -- 3.34 & 2.50 -- 3.54 \\
\hline%---------------------------------------------------------------------
$|\Delta m^2|/10^{-3}~\mathrm{eV}^2 $ & NO  & 2.525 & 2.495 -- 2.567 & 2.454 -- 2.606 & 2.411 -- 2.646 \\
                                      & IO  & 2.505 & 2.473 -- 2.539 & 2.430 -- 2.582 & 2.390 -- 2.624 \\
                                      & Any & 2.525 & 2.495 -- 2.567 & 2.454 -- 2.606 & 2.411 -- 2.646 \\
\hline%---------------------------------------------------------------------
$\sin^2 \theta_{13}/10^{-2}$ & NO & 2.15 & 2.08 -- 2.22 & 1.99 -- 2.31 & 1.90 -- 2.40 \\
                             & IO & 2.16 & 2.07 -- 2.24 & 1.98 -- 2.33 & 1.90 -- 2.42 \\
                             & Any & 2.15 & 2.08 -- 2.22 & 1.99 -- 2.31 & 1.90 -- 2.40 \\
\hline%---------------------------------------------------------------------
$\sin^2 \theta_{23}/10^{-1}$ & NO & 4.25 & 4.10 -- 4.46 & 3.95 -- 4.70 & 3.81 -- 6.15 \\
                             & IO & 5.89 & 4.17 -- 4.48 $\oplus$ 5.67 -- 6.05 & 3.99 -- 4.83 $\oplus$ 5.33 -- 6.21 &  3.84 -- 6.36 \\
                             & Any & 4.25 & 4.10 -- 4.46  & 3.95 -- 4.70 $\oplus$ 5.75 -- 6.00 & 3.81 -- 6.26 \\
\hline%---------------------------------------------------------------------
$\delta/\pi$ & NO & 1.38 & 1.18 -- 1.61 & 1.00 -- 1.90  &  0 -- 0.17 $\oplus$ 0.76 -- 2   \\
             & IO & 1.31 & 1.12 -- 1.62 & 0.92 -- 1.88  &  0 -- 0.15 $\oplus$ 0.69 -- 2  \\
             & Any & 1.38 & 1.18 -- 1.61 & 1.00 -- 1.90  &  0 -- 0.17 $\oplus$ 0.76 -- 2  \\
%\hline%---------------------------------------------------------------------
%$\Delta \chi^2_{\mathrm{{IO}-{NO}}}$ & IO$-$NO & +3.6  \\ [1pt]
%\hline\hline
\end{tabular}
\end{ruledtabular}
%}%end of resizebox
\vspace*{-.4cm}
\end{table}
%============================================================================

The parameter $\sin^2\theta_{23}$ is less well known, at the level of 9.6\%.  At $\!3\sigma$, its
octant degeneracy is unresolved, and maximal mixing is also allowed. 
At lower significance, maximal mixing is disfavored in both NO and IO,
and the first octant is preferred in NO. The $n\sigma$ ranges for $\theta_{23}$ for any ordering
are larger than for NO (Table~I), as a result of joining 
the NO and IO intervals determined by the curves in the right-lower panel of Fig.~1 at $\chi^2=n^2$.
   Concerning the possible CP-violating phase $\delta$, 
our analysis strengthen the trend in favor of $\delta \sim 3\pi/2$ \cite{Capo16,Conc17,T2K2017}, and disfavors 
ranges close $\delta \sim \pi/2$ at $\gtrsim 3\sigma$. 
In any case, the parameters $\theta_{23}$ and $\delta$ do not enter in the calculation of 
($m_\beta,\,m_{\beta\beta},\,\Sigma$).  

A few remarks are in order about the IO-NO offset in Eq.~(\ref{preferNO}).  This value is in
the ballpark of the official SK fit results quoted
in \cite{SKatm1,SKatm2}, namely: 
$\Delta\chi^2_\mathrm{IO-NO} = 4.3$ (for SK data at fixed $\theta_{13}$)
and $\Delta\chi^2_\mathrm{IO-NO} = 5.2$ (for SK + T2K data at fixed $\theta_{13}$). 
By excluding SK atmospheric data in our fit, we find $\Delta\chi^2_\mathrm{IO-NO} = 1.1$, 
in qualitative accord with the official T2K data analysis constrained by reactor data 
\cite{T2K2017}.

Concerning SK atmospheric data, it has been emphasized \cite{Capo16,Conc17,Vall14} that
the complete set of bins and systematics \cite{SKatm1,SKatm2} can only be handled  
within the collaboration, especially when $\nu/\overline\nu$ or multi-ring 
event features are involved. Nevertheless, 
we think it useful to continue updating our analysis of 
reproducible SK samples, namely, sub/multi-GeV 
single-ring ($e$-like and $\mu$-like) and stopping/through-going ($\mu$-like) distributions. 
These samples encode interesting (although entangled and smeared) pieces of
information about subleading effects 
driven by known and unknown oscillation parameters, see e.g.\ \cite{Glob06}; 
in particular, they contributed to early hints of nonzero $\theta_{13}$ \cite{Hint}.
At present, we trace the atmospheric hint of NO to $e$-like events, especially multi-GeV, in 
qualitative agreement with  \cite{Kearns}.%
%-----------------
\footnote{Note, however, that weaker results for the IO-NO difference ($\lesssim 1\sigma$), 
with or without atmospheric data, have been found in  \protect\cite{Conc17}.}
%--------------------

Summarizing, the SK(+T2K) official results in \cite{SKatm1,SKatm2,T2K2017} and ours in
Eq.~(\ref{preferNO}) suggest, at face value, that
global $3\nu$ oscillation analyses may have reached an overall $\sim\! 2\sigma$ sensitivity
to the mass ordering, with a preference for NO driven by atmospheric data and
corroborated by accelerator data, together with reactor constraints.
This intriguing indication, although still tentative, is generally supported by cosmological 
data (see Sec.~II~C) and thus
warrants a dedicated discussion in the context of 
absolute $\nu$ mass observables (see Sec.~III).

\subsection{Neutrinoless double beta decay}

If the three known neutrinos are Majorana fermions, the rare process of $0\nu\beta\beta$ decay is expected to occur
with half life $T$ given by
%-----------
\begin{equation}
\label{decay}
T^{-1} = G\, |M|^2\, m^2_{\beta\beta}\ ,
\end{equation}
%---------------
where $m_{\beta\beta}$ is given in Eq.~(\ref{mbb}), 
$G$ is the (calculable) phase space and $M$ is the nuclear
matrix element (NME) for a candidate nucleus \cite{Viss16,Giun15,Paes15,Verg16}. 

A worldwide search is underway to find possible $0\nu\beta\beta$ decay
signatures in a variety of nuclei, and lower limits on the corresponding half lives have been placed 
\cite{Schw13,Crem14,Bara17}. 
Transforming lower bounds on $T$ into upper bounds on $m_{\beta\beta}$ requires theoretical input on the  NME
and their uncertainties \cite{Viss16,Giun15,Paes15,Verg16,Voge12,Suho15,Enge17}. 
The strongest $m_{\beta\beta}$ limit to date is provided by the KamLAND-Zen experiment with  ${}^{136}$Xe, that
finds $T > 1.07 \times 10^{26}$ yr (90\% C.L.), and derives the range
$m_{\beta\beta}\gtrsim 0.061$--0.165~eV (90\% C.L.) by bracketing recent NME calculations \cite{KamZ16}. For the sake
of simplicity, we include only the (dominant) KamLAND-Zen constraints herein.

In order to derive bounds at any given C.L.\ in our analysis, we build a general $\chi^2(m_{\beta\beta})$ function
by using: $(i)$ the experimental $\chi^2(T)$ curve presented by the KamLAND-Zen collaboration in \cite{KLchi2} [with $T=T(m_{\beta\beta},\, |M|)$ 
from Eq.~(\ref{decay})]; and $(ii)$ our construction of the $\chi^2(|M|)$ function, based on the conservative theoretical
uncertainties estimated in \cite{Rotu15}. The objective function is obtained as 
\begin{equation}
\label{obj}
\chi^2(m_{\beta\beta}) = \min_{|M|} \left[ \chi^2\left(T(m_{\beta\beta},\, |M|)\right) + \chi^2(|M|)\right]\ ,
\end{equation}
and is shown in Fig.~2. These results do not depend on the mass ordering and, in particular, 
$\Delta\chi^2_\mathrm{IO-NO}=0$.

%%%%%%%%%%%%%%%%%%%%%%%%%%%%%%%%%%%%%%%%%%%%%%%%%%%%%%%%%%%%%%%%%%%%%%%%%%%%%%%%%%%%%%%%%%
\begin{figure}[t]
\begin{minipage}[c]{0.96\textwidth}
\includegraphics[width=0.58\textwidth]{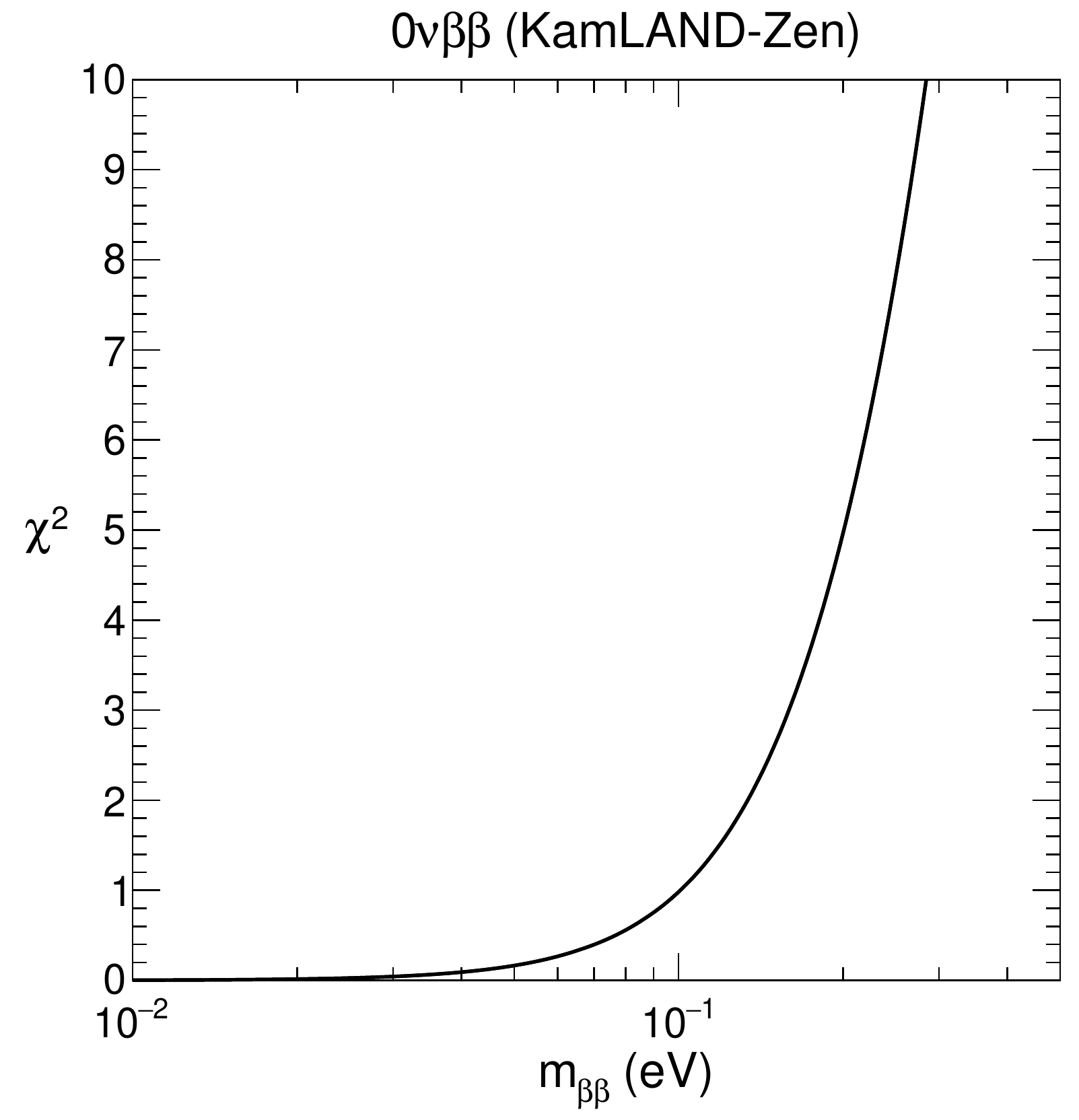}
\caption{\label{fig02}\footnotesize  Constraints from $0\nu\beta\beta$ decay, in terms of the function $\chi^2(m_{\beta\beta})$ 
derived from KamLAND-Zen data \protect\cite{KamZ16,KLchi2} 
and from an estimate of the ${}^{136}$Xe nuclear matrix elements and its uncertainties based
on \protect\cite{Rotu15}. The same constraints apply to both NO and IO.  See the text for details.} 
\end{minipage}
\end{figure}
%%%%%%%%%%%%%%%%%%%%%%%%%%%%%%%%%%%%%%%%%%%%%%%%%%%%%%%%%%%%%%%%%%%%%%%%%%%%%%%%%%%%%%%%%%

From Fig.~2 we get $m_{\beta\beta}<0.15$~eV at 90\% C.L., close to the 
most conservative limit quoted at the same C.L.\ in \cite{KamZ16} ($0.165$~eV). From Fig.~2 we also derive
%---------------
\begin{equation}
\label{mbb-bounds}
m_{\beta\beta}< 0.18\ \mathrm{eV\ at}\ 2\sigma\ (<0.27 \ \mathrm{eV\ at}\ 3\sigma)\ .
\end{equation} 
%------------------
For completeness, our assessment of the $\chi^2(m_{\beta\beta})$ function is detailed below. 

According to \cite{Rotu15}, Eq.~(\ref{decay}) is linearized via logarithms as
$\tau = \gamma - 2\eta - 2\mu$,
where: $\gamma = -\log_{10}(G/\mathrm{y}^{-1}\mathrm{eV}^{-2}$), 
$\tau= \log_{10}(T/y)$,  $\eta = \log_{10}(|M|)$,  and $\mu=\log_{10}(m_{\beta\beta})$.  The NME $\eta$ 
and its uncertainties with respect to a central value $\overline\eta$ are parametrized as 
$ \eta=\overline\eta  + \alpha (g_A-1) + s\beta \pm \sigma $,
where $g_A$ is the effective axial coupling (typically 
``quenched'' with respect to the vacuum value $g_A^0 \simeq 1.27$ \cite{Viss16}), $s=\pm 1$ switches between two alternative approaches
to short-range correlation effects (so-called CD-Bonn and Argonne potentials), while
$\sigma$ is a residual (nonparametric) uncertainty \cite{Rotu15}. Numerical values for 
${}^{136}$Xe are: $\gamma=24.865$, $\overline\eta = 0.267$, $\alpha=0.458$, $\beta=0.021$, $\sigma=0.032$.

Concerning the total $1\sigma$ uncertainty $\sigma_\eta$ affecting $\eta$, we assume $g_A=1\pm 0.15$ as a reasonable
$1\sigma$ estimate for the axial coupling. The central value corresponds to default quenching $(g_A=1)$ \cite{Viss16}, and 
 the $2\sigma$ range $g_A\in [0.7,\,1.3]$ spans typical effective values adopted
in the NME literature up to $g_A^0$ \cite{Verg16}, while the $3\sigma$ range goes down to $g_A=0.55$, 
close to the very low estimates $g_A \simeq  1.27 A^{-0.18} \simeq 0.52$  considered in \cite{Viss16,Bare15} (for $A=136$).  
Concerning alternative short-range correlation approaches, we conservatively assume that 
the associated uncertainty  ($s\beta = \pm\beta$) corresponds statistically to $\pm 1\sigma$.   
The total error $\sigma_\eta$ is then evaluated by summing in quadrature the three independent components, 
namely, $\alpha \cdot 0.15 = 0.069$,
$\beta=0.021$ and $\sigma=0.032$, leading to 
$\eta = \overline\eta \pm \sigma_\eta = 0.267 \pm 0.079$ $(1\sigma)$.
Finally,
we minimize over $\eta$ according to Eq.~(\ref{obj}), where the second term on the right-hand side 
is given by $\chi^2_\eta = [(\eta-\overline\eta)/\sigma_\eta]^2$.   
Our estimate $\eta = 0.267 \pm 0.079$ implies a $\pm3\sigma$ range $|M|\simeq 1.1$--3.2, to be compared with the 
total range $|M|\simeq 1.6$--4.3 adopted in \cite{KamZ16}. The overall shift is mainly related to a different choice for the default 
axial coupling ($g_A=1.27$ in \cite{KamZ16} versus $g_A=1$ herein). In any case, both ranges correspond to a conservative
factor of $\sim\!3$ uncertainty of $|M|$. 
 
The above results  
refer to a single (dominant) experimental datum for the $^{136}$Xe nucleus. When comparable bounds on $m_{\beta\beta}$
will be obtained in other experiments and nuclei, the combination of various $0\nu\beta\beta$ data  
should  take into account the theoretical NME covariances among different nuclei \cite{QRPA08}.

\subsection{Cosmology}

Neutrinos are the only known particles  in the standard model of particle physics that can change behaviour, from the relativistic to
the non-relativistic regime, in an epoch after cosmic microwave background (CMB) recombination. This change leaves a characteristic imprint on several
cosmological observables (see, e.g. \cite{Hann10, Hu:1997mj, Dolgov:2002wy, Lesgourgues:2006nd, Abazajian:2011dt, Lesgourgues:2012uu}), 
letting cosmology to strongly bound the neutrino mass scale, indeed providing the current strongest (albeit model dependent) bounds on $\Sigma$.
Bounds on $\Sigma$ from recent cosmological data have been presented in several papers (see, for example, \cite{Vagnozzi:2017ovm,Giusarma:2016phn,Xu:2016ddc, Huang:2015wrx, DiValentino:2015sam, Cuesta:2015iho} and references therein) while forecasts for near (and far) future cosmological datasets have been obtained in
\cite{Archidiacono:2016lnv, Oyama:2015gma, Allison:2015qca, DiValentino:2016foa, Banerjee:2016suz, Hamann:2012fe}.

Clearly, current cosmological constraints on neutrino masses depend on the combination of datasets considered and on the
theoretical framework assumed (see, for example, \cite{DiValentino:2015ola, DiValentino:2016hlg, Motohashi:2012wc, Hu:2014sea, Wang:2016tsz, Zhang:2015uhk}). 
It is therefore important to be extremely clear in the description of the assumptions we make. In our analysis we consider $6$ different combinations of the following datasets:

\begin{itemize}

\item The full range of the Planck 2015 temperature anisotropy angular power spectrum, both at low multipole $\ell$ ($2\le \ell \le 29)$ and high $\ell$ ($30\le \ell \le 2508)$, provided by the Planck collaboration \cite{Aghanim:2015xee}. We define this dataset as Planck {\scriptsize TT}.

\item The full multipole range of the Planck 2015 temperature anisotropy angular power spectrum, and high multipoles E polarization and cross TE temperature polarization anisotropy angular power spectra ($30\le \ell \le 2508)$) \cite{Aghanim:2015xee}. We define this dataset as Planck {\scriptsize TT, TE, EE}.

\item A gaussian prior on the reionization optical depth $\tau=0.055\pm0.009$, as obtained recently from Planck HFI data \cite{newtau}. We refer to this prior
as $\tau_\mathrm{HFI}$.

\item The Baryon Acoustic Oscillation measurements from 6dFGS \cite{beutler2011}, SDSS-MGS \cite{ross2014}, BOSSLOWZ \cite{anderson2014} and CMASS-DR11 \cite{anderson2014} surveys as done in \cite{planckparams2015}. We label this dataset as {BAO}.

\item The Planck 2015 CMB lensing potential power spectrum reconstruction data  \cite{Ade:2015zua}. We refer to this dataset as {``lensing''}.

\end{itemize}

In our analysis we always consider a flat universe, a cosmological constant and adiabatic primordial perturbations, within the so-called $\Lambda$CDM model.
However, we consider two slightly different theoretical scenarios: 

\begin{itemize}

\item The ``standard'' $6+1$ parameters of the $\Lambda$CDM + $\Sigma$ model, where the $6$ parameters of the $\Lambda$CDM are the
baryon and Cold Dark Matter densities $\omega_b$ and $\omega_{cm}$, the amplitude $A_s$ and spectral index $n_s$  of primordial density fluctuations,
the Hubble constant $H_0$ and the reionization optical depth $\tau$, and the seventh extra parameter $\Sigma$ is also free; 

\item An extended $6+2$ parameter scenario, considering variation also in the lensing amplitude $A_\mathrm{lens}$ that controls the effects of gravitational lensing in the
Planck TT, TE and EE angular spectra \cite{cala}. This parameter is expected to be $A_\mathrm{lens}=1$ in the standard $\Lambda$CDM model. However the most recent Planck data analysis shows a statistically significant preference for values $A_\mathrm{lens}>1$ (in particular, $A_\mathrm{lens}=1.15_{-0.12}^{+0.13}$ at $\pm 2\sigma$) \cite{newtau}. While the physical motivations behind this result are not yet clear (systematics or new physics) we consider also this parameter as free, since its correlation with $\Sigma$ strongly weakens the cosmological constraints on neutrino masses. The scenario with extra $A_\mathrm{lens}$ parameter is therefore expected to yield more conservative results.
\end{itemize}

%===========================================================================
\begin{table}[t]
%\centering
%\captionsetup{width=.96\textwidth}
\caption{\label{Cosmology} \footnotesize  Results of the global $3\nu$ analysis of cosmological data 
within the standard $\Lambda\mathrm{CDM}+\Sigma$   and extended $\Lambda\mathrm{CDM}+\Sigma+A_\mathrm{lens}$ models. The datasets refer to various combinations
of the Planck power
angular CMB temperature power spectrum (TT) 
plus polarization power spectra (TE, EE), reionization optical depth $\tau_\mathrm{HFI}$, lensing potential power spectrum (lensing), and BAO
measurements.
For each of the 12 cases we report the $2\sigma$ upper bounds on $\Sigma=m_1+m_2+m_3$ for NO and IO, together
with the $\Delta\chi^2$ difference between the two mass orderings (with one digit after decimal point). For any $\Sigma$, 
the masses $m_i$ are taken 
to obey the $\delta m^2$ and $\Delta m^2$ constraints coming from oscillation data. See the text for more details. }
\vspace*{0mm}
\centering
\begin{ruledtabular}
\begin{tabular}{cllccc}
\# & Model & Cosmological data set & $\Sigma/$eV ($2\sigma$), NO  & $\Sigma/$eV ($2\sigma$), IO & $\Delta\chi^2_\mathrm{IO-NO}$   \\[1mm]
\hline%---------------------------------------------------------------------
 1 & $\Lambda\mathrm{CDM}+\Sigma$ 				& Planck {\scriptsize TT} + $\tau_\mathrm{HFI}$   					& $<0.72$ & $<0.80$ & $ 0.7$   \\
 2 & $\Lambda\mathrm{CDM}+\Sigma$					& Planck {\scriptsize TT} + $\tau_\mathrm{HFI}$ + lensing			& $<0.64$ & $<0.63$ & $ 0.2$   \\
 3 & $\Lambda\mathrm{CDM}+\Sigma$ 				& Planck {\scriptsize TT} + $\tau_\mathrm{HFI}$ + BAO				& $<0.21$ & $<0.23$ & $ 1.2$   \\
\hline%---------------------------------------------------------------------
 4 & $\Lambda\mathrm{CDM}+\Sigma$					& Planck {\scriptsize TT,\,TE,\,EE} + $\tau_\mathrm{HFI}$			& $<0.44$ & $<0.48$ & $ 0.6$   \\
 5 & $\Lambda\mathrm{CDM}+\Sigma$ 				& Planck {\scriptsize TT,\,TE,\,EE} + $\tau_\mathrm{HFI}$ + lensing	& $<0.45$ & $<0.47$ & $ 0.3$   \\
 6 & $\Lambda\mathrm{CDM}+\Sigma$ 				& Planck {\scriptsize TT,\,TE,\,EE} + $\tau_\mathrm{HFI}$ + BAO		& $<0.18$ & $<0.20$ & $ 1.6$   \\
\hline%---------------------------------------------------------------------
 7 & $\Lambda\mathrm{CDM}+\Sigma+A_\mathrm{lens}$ 	& Planck {\scriptsize TT} + $\tau_\mathrm{HFI}$  					& $<1.08$ & $<1.08$ & $-0.1$\phantom{+}   \\
 8 & $\Lambda\mathrm{CDM}+\Sigma+A_\mathrm{lens}$ 	& Planck {\scriptsize TT} + $\tau_\mathrm{HFI}$ + lensing			& $<0.91$ & $<0.93$ & $ 0.0$  \\
 9 & $\Lambda\mathrm{CDM}+\Sigma+A_\mathrm{lens}$	& Planck {\scriptsize TT} + $\tau_\mathrm{HFI}$ + BAO				& $<0.45$ & $<0.46$ & $ 0.2$   \\
\hline%---------------------------------------------------------------------
10 & $\Lambda\mathrm{CDM}+\Sigma+A_\mathrm{lens}$	& Planck {\scriptsize TT,\,TE,\,EE} + $\tau_\mathrm{HFI}$ 			& $<1.04$ & $<1.03$ & $ 0.0$   \\
11 & $\Lambda\mathrm{CDM}+\Sigma+A_\mathrm{lens}$	& Planck {\scriptsize TT,\,TE,\,EE} + $\tau_\mathrm{HFI}$ + lensing	& $<0.89$ & $<0.89$ & $ 0.1$   \\
12 & $\Lambda\mathrm{CDM}+\Sigma+A_\mathrm{lens}$	& Planck {\scriptsize TT,\,TE,\,EE} + $\tau_\mathrm{HFI}$ + BAO		& $<0.31$ & $<0.32$ & $ 0.3$   \\
\end{tabular}
\end{ruledtabular}
%\vspace*{.6cm}
\end{table}
%============================================================================

The cosmological constraints are obtained using the November 2016 version of the publicly available Monte-Carlo Markov Chain package \texttt{cosmomc} \cite{Lewis:2002ah, Lewis:1999bs}, with a convergence diagnostic based on the Gelman and Rubin statistic, that implements an efficient sampling of the posterior distribution using the fast/slow parameter decorrelations \cite{Lewis:2013hha}, and that includes the support for the Planck data release 2015 Likelihood Code \cite{Aghanim:2015xee} (see \url{http://cosmologist.info/cosmomc/}). We emphasize that we implement separately the NO and IO options in the CosmoMC analysis, namely,
the masses $m_i$ entering in the definition of $\Sigma$ obey the
$\delta m^2$ and $\Delta m^2$ constraints in Eqs.~(\ref{delta1},\ref{delta2}). In particular,
from the fit results in Table~I and Eq.~(\ref{lightest1}), it is  
%...................
\begin{equation}
\label{lightest2}
{\Sigma} = m_1+m_2+m_3 \gtrsim \left\{
\begin{array}{lc}
0.06\ \mathrm{eV} & \rm{(NO)}\ ,\\
0.10\ \mathrm{eV} & \rm{(IO)}\ .\\
\end{array}\right.
\end{equation}
%....................
Such approach differs from other recent studies, where neutrino masses are assumed to be degenerate ($m_i = m\geq 0$) and the above constraints
are relaxed ($\Sigma\geq 0$). In such studies, best-fit results around $\Sigma \simeq 0$ (i.e., in the unphysical region) tend to
induce a somewhat artificial preference for NO over IO, just because NO allows  $\Sigma$ values lower than IO.   
For any scenario and combination of cosmological datasets, our CosmoMC fit leads, in general, to 
different best-fit values for $\Sigma$ (in the physical region) 
and for the associated values of $\chi^2_{\min}$ in NO and IO.  The value of $\Delta \chi^2_\mathrm{IO-NO}$ 
correctly quantifies the overall preference of the fitted cosmological data set for one mass ordering. 

From CosmoMC one also gets the
posterior  probability functions $p(\Sigma)$ in NO and IO, which are transformed into $\chi^2(\Sigma)$ 
functions by applying \cite{Stat16} the standard Neyman construction \cite{Neym67} and Feldman-Cousins method \cite{Feld98}.
We have also verified that, for any given cosmological data set, the resulting $\chi^2(\Sigma)$ curves for NO and IO 
converge for increasing $\Sigma$ as they should (up to residual numerical artifacts at the level of $\delta\chi^2\lesssim 0.1$).  
The $\chi^2$ analysis of cosmological data is thus methodologically consistent with the $\chi^2$ analysis of oscillation
and $0\nu\beta\beta$ data, and a global combination of the data can  be performed (see next Section).

The main cosmological fit results are summarized in Table~\ref{Cosmology}, in terms of upper bounds (at $2\sigma$ level)
on the sum of neutrino masses $\Sigma$ for NO and IO, together with the $\Delta\chi^2_\mathrm{NO-IO}$ offset.
The results show some global trends: (a) the $\Sigma$ bounds are significantly strengthened by enlarging the Planck temperature data 
with polarization spectra or with BAO data, while they are only moderately tightened by adding lensing data; (b) the bounds are
largely weakened, up to a factor of $\sim\!2$ in some cases, by letting $A_\mathrm{lens}$ free. 

%%%%%%%%%%%%%%%%%%%%%%%%%%%%%%%%%%%%%%%%%%%%%%%%%%%%%%%%%%%%%%%%%%%%%%%%%%%%%%%%%%%%%%%%%%
\begin{figure}[ht]
\begin{minipage}[c]{0.96\textwidth}
\includegraphics[width=0.44\textwidth]{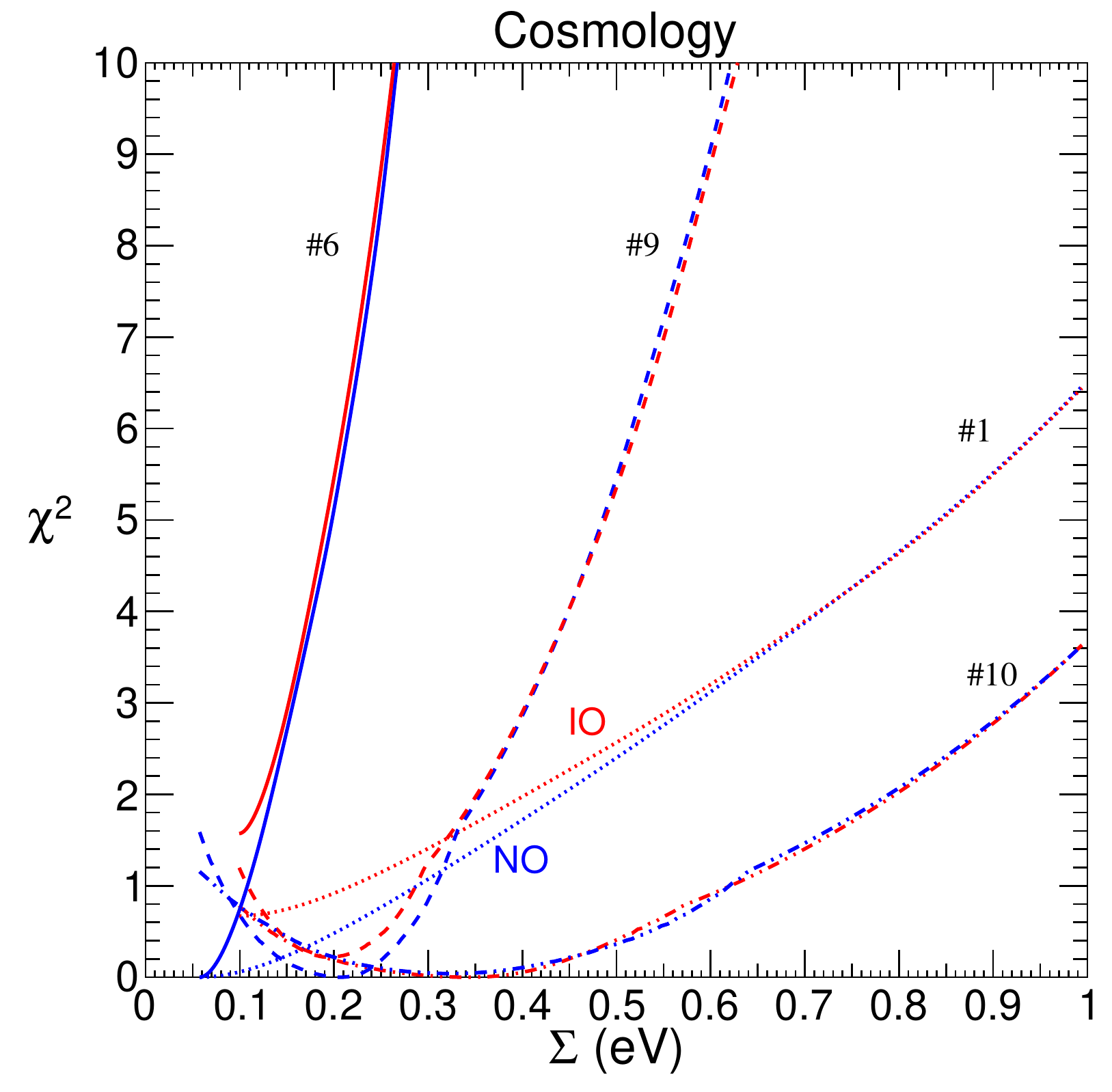}
\caption{\label{fig03}\footnotesize Constraints on the sum of neutrino masses from cosmological data. The $\chi^2(\Sigma)$ function is shown in NO (blue) and IO (red) for four
representative cases, numbered as \#10, \#1, \#9, and \#6 in Table~II, and  
including the corresponding $\Delta\chi^2_\mathrm{IO-NO}$ offset. In each case, 
the NO and IO curves diverge as $\Sigma$ approaches the extrema 
in Eq.~(\protect\ref{lightest2}), while they tend to converge for large $\Sigma$, as the mass
ordering sensitivity vanishes.} 
\end{minipage}
\end{figure}
%%%%%%%%%%%%%%%%%%%%%%%%%%%%%%%%%%%%%%%%%%%%%%%%%%%%%%%%%%%%%%%%%%%%%%%%%%%%%%%%%%%%%%%%%%

At a finer level, slight differences emerge between the results in NO and IO in Table~II, indicating 
a weak sensitivity of cosmological data to the mass ordering. Interestingly, 
normal ordering is generally preferred, except for a few cases where $\Delta\chi^2_\mathrm{NO-IO}$
is either negligible or slightly negative, corresponding to the extended and conservative scenario in which the 
$A_\mathrm{lens}$ parameter is varying. The overall indication in favor of NO, although still
at the $\lesssim 1\sigma$ level, is consistent with the neutrino
oscillation results in Eq.~(\ref{preferNO}), and brings the global preference for NO
at the typical level of $\Delta\chi^2 \simeq 4$ in our analysis (i.e., $2\sigma$).
Note that the overall preference for NO from cosmological 
data exceeds $1\sigma$ only in the cases \#3 and \#6 of Table~\ref{Cosmology} that, not surprisingly, are 
associated with the strongest constraints on the sum of neutrino masses ($\Sigma\lesssim 0.2$~eV at $2\sigma$), that are arising when using the BAO data, since they are directly sensitive to the free-streaming nature of the neutrinos. Moreover, the constraints of these two cases are not affected by the lowering of the optical depth, as it happens for the other combination of datasets if compared to the Planck 2015 findings \cite{planckparams2015}, showing that the BAO bounds are very robust and reliable.

Further details can be appreciated in terms of the $\chi^2(\Sigma)$ functions for NO and IO. For the sake of 
simplicity, we do so only for four representative cases numbered as \#10, \#1, \#9, and \#6 in Table~\ref{Cosmology},
as  shown in Fig.~\ref{fig03}.  
These cases lead to increasingly strong upper bounds on $\Sigma$, ranging
from $\lesssim 1$~eV (\#10) to $\lesssim 0.2$~eV  (\#6) at $2\sigma$. The corresponding
offset $\Delta\chi^2_\mathrm{IO-NO}$ ranges from $\sim 0$ to $1.6$. 
In two cases
(\#1 and \#6)  the $\chi^2$ minima are reached at the extrema of $\Sigma$ 
from Eq.~(\ref{lightest2}), while in the other two (\#10 and \#9) they are reached
at higher values. In general, we find that  the cases with $A_\mathrm{lens}$ free lead to
best-fit values  of $\Sigma$ above the extrema in Eq.~(\ref{lightest2}) (not shown). 
We emphasize that, in all cases,  the $\chi^2(\Sigma)$ curves for NO and IO tend to converge for 
degenerate masses $m_i$ (large $\Sigma$), while they bifurcate towards the extrema in Eq.~(\ref{lightest2}), 
corresponding to strongly hierarchical masses at low $\Sigma$. 
The four cases in Fig.~\ref{fig03} are sufficiently representative
of the variety of constraints set by current cosmological data in Table~\ref{Cosmology}, 
and will be explicitly considered in the global analysis of oscillation and non-oscillation neutrino data in the next Section.
   
\section{Combined constraints in the $\bm{(\Sigma,\,m_{\beta\beta})}$ plane}

In this Section we present increasingly strong constraints on the absolute mass observables 
${(\Sigma,\,m_{\beta\beta})}$ in the (sub)eV range, obtained by combining the $\chi^2$ from oscillation data (Fig.~1) with the $\chi^2$ from $0\nu\beta\beta$ (Fig.~2) and then with the $\chi^2$ from cosmological data (Fig.~3). Current $\beta$-decay constraints 
($m_\beta\lesssim 2$~eV \cite{PDG2016})
are not relevant in this context. 

As discussed in Sec.~II, we consider two alternative ways to obtain
allowed regions: $(i)$ the $\chi^2$ is separately minimized on 
the relevant parameters in each mass ordering, either NO or IO,
discarding the $\Delta \chi^2_\mathrm{IO-NO}$ information; and $(ii)$ the $\chi^2$ 
is further minimized over NO and IO, including the $\Delta \chi^2_\mathrm{IO-NO}$ information. 
In the former case, one should consider the NO and IO allowed regions as exclusive while, in the latter case, one should join the NO and IO allowed regions to obtain the global ones in ``any ordering''.

%%%%%%%%%%%%%%%%%%%%%%%%%%%%%%%%%%%%%%%%%%%%%%%%%%%%%%%%%%%%%%%%%%%%%%%%%%%%%%%%%%%%%%%%%%
\begin{figure}[t]
\begin{minipage}[c]{0.96\textwidth}
\includegraphics[width=0.70\textwidth]{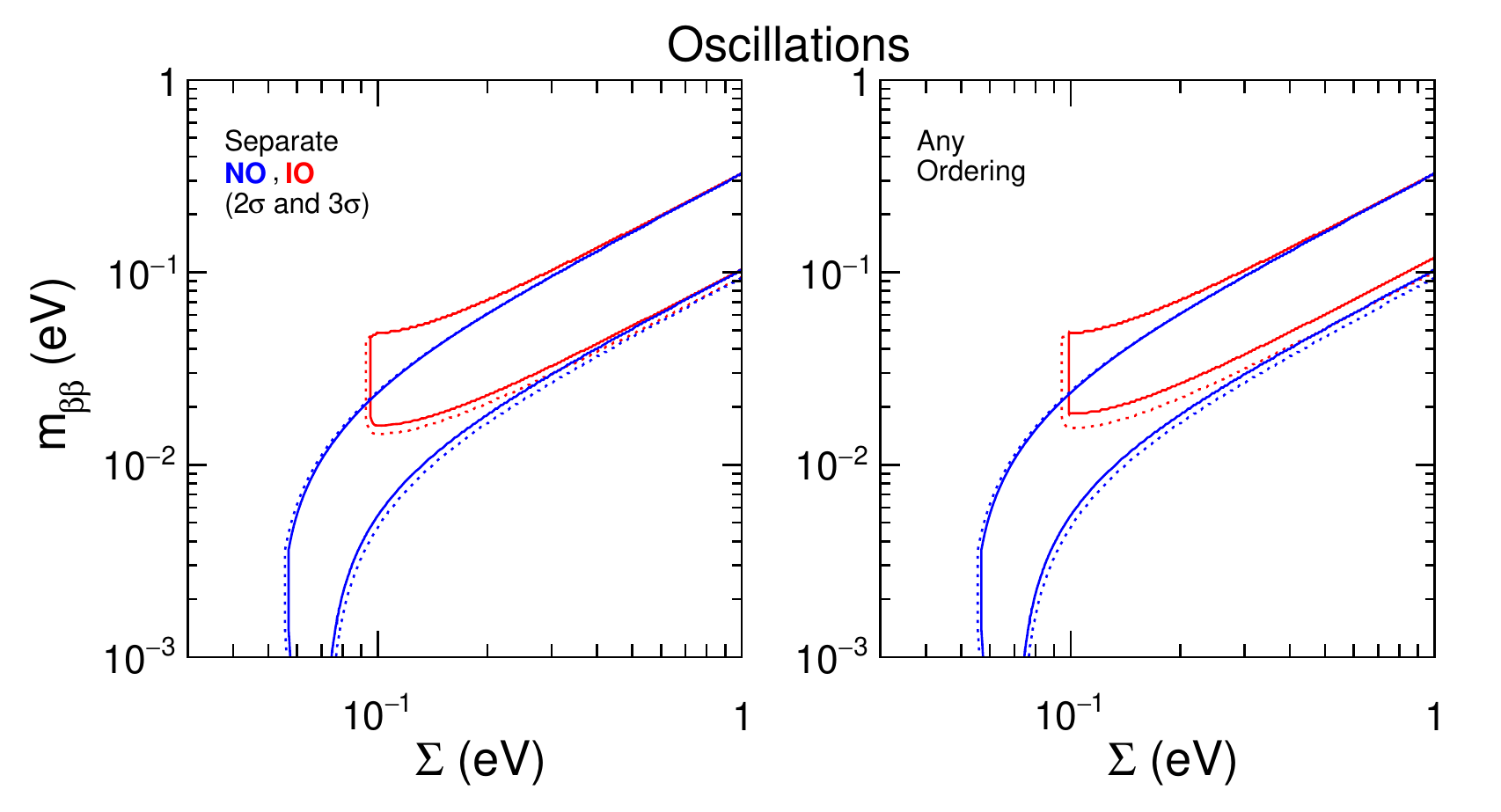}
\caption{\label{fig04}\footnotesize Global analysis in the $(\Sigma,\,m_{\beta\beta})$ plane, including only oscillation data. 
Constraints are shown in terms of $2\sigma$ (solid) and 
$3\sigma$ (dotted) allowed regions for NO (blue) and IO (red). In the left panel,
the $\chi^2$ minimization is separately performed in each mass ordering, and the
allowed regions should be separately considered for NO and IO. In the right
panel, the $\chi^2$ is further minimized over the mass ordering, and the allowed
regions (for any ordering) are given by the union of the NO and IO ones. 
} 
\end{minipage}
\end{figure}
%%%%%%%%%%%%%%%%%%%%%%%%%%%%%%%%%%%%%%%%%%%%%%%%%%%%%%%%%%%%%%%%%%%%%%%%%%%%%%%%%%%%%%%%%%

%%%%%%%%%%%%%%%%%%%%%%%%%%%%%%%%%%%%%%%%%%%%%%%%%%%%%%%%%%%%%%%%%%%%%%%%%%%%%%%%%%%%%%%%%%
\begin{figure}[th]
\begin{minipage}[c]{0.96\textwidth}
\includegraphics[width=0.70\textwidth]{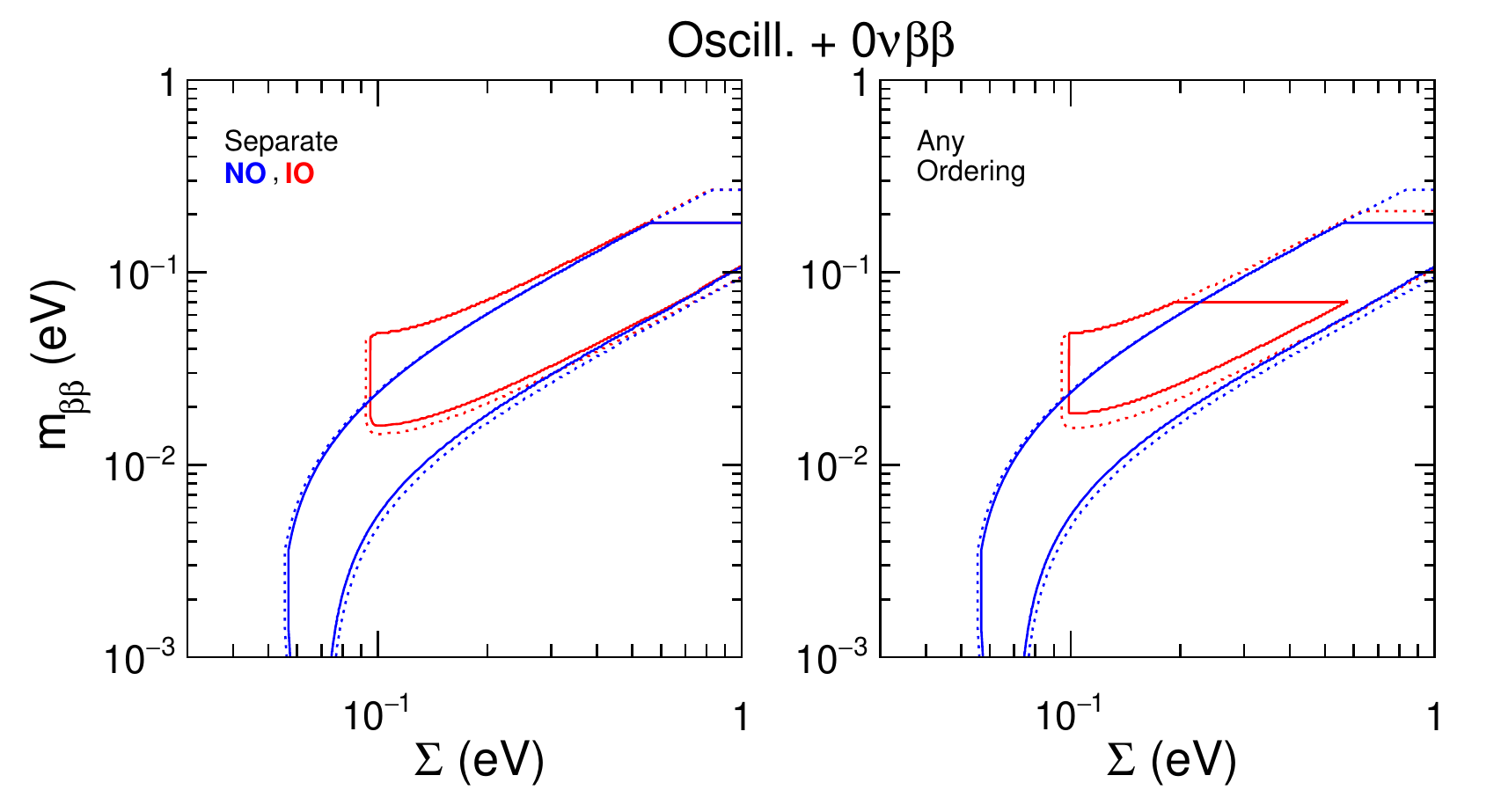}
\caption{\label{fig05}\footnotesize As in Fig.~4, but including the $\chi^2({m_{\beta\beta}})$
function from Fig.~2.} 
\end{minipage}
\end{figure}
%%%%%%%%%%%%%%%%%%%%%%%%%%%%%%%%%%%%%%%%%%%%%%%%%%%%%%%%%%%%%%%%%%%%%%%%%%%%%%%%%%%%%%%%%%

Figure~4 shows the $2\sigma$ 
and $3\sigma$ constraints in the ${(\Sigma,\,m_{\beta\beta})}$ plane, 
derived from the
oscillation data discussed in Sec.~II~A. The left panel refers to separate
fits in each mass ordering, while the right panel to the global fit in any ordering.   
The main features of the allowed bands have been discussed in previous literature 
(see \cite{Melc04,Melc06,Rodejo1,DellOro1,DellOro2,Zavanin} and refs.\ therein)
and are not repeated here. We only recall that the vertical width of the bands is mainly
related to the unknown Majorana phases, while the oscillation parameter uncertainties play a 
secondary role, that can be appreciated via the difference between the $2\sigma$ and
$3\sigma$ allowed regions.  By comparing the left and right panels in Fig.~4, one can notice
that the NO regions are identical, while the IO region is slightly reduced on the right,
due to the offset of the $\chi^2$ minimum for IO in Eq.~(\ref{preferNO}). 

Figure~5 is similar to Fig.~3, but includes the $0\nu\beta\beta$ constraints 
discussed in Sec.~II~B.  In the left panel, both the NO and the IO
allowed bands are horizontally cut at the $m_{\beta\beta}$ values in Eq.~(\ref{mbb-bounds}).
In the right panel, the upper bounds on $m_{\beta\beta}$ are stronger in IO, 
and can be estimated by drawing in Fig.~2 the 
lines at $\chi^2= n^2 - 3.6$ ($n=2,\,3$), where 3.6 corresponds to the offset in Eq.~(\ref{preferNO}). 
Note that, in the left panel, the projections of the NO and IO allowed regions 
onto the abscissa lead to upper bounds on $\Sigma$ well above 1~eV.

\newpage

%%%%%%%%%%%%%%%%%%%%%%%%%%%%%%%%%%%%%%%%%%%%%%%%%%%%%%%%%%%%%%%%%%%%%%%%%%%%%%%%%%%%%%%%%%
\begin{figure}[t]
\begin{minipage}[c]{0.96\textwidth}
\includegraphics[width=0.70\textwidth]{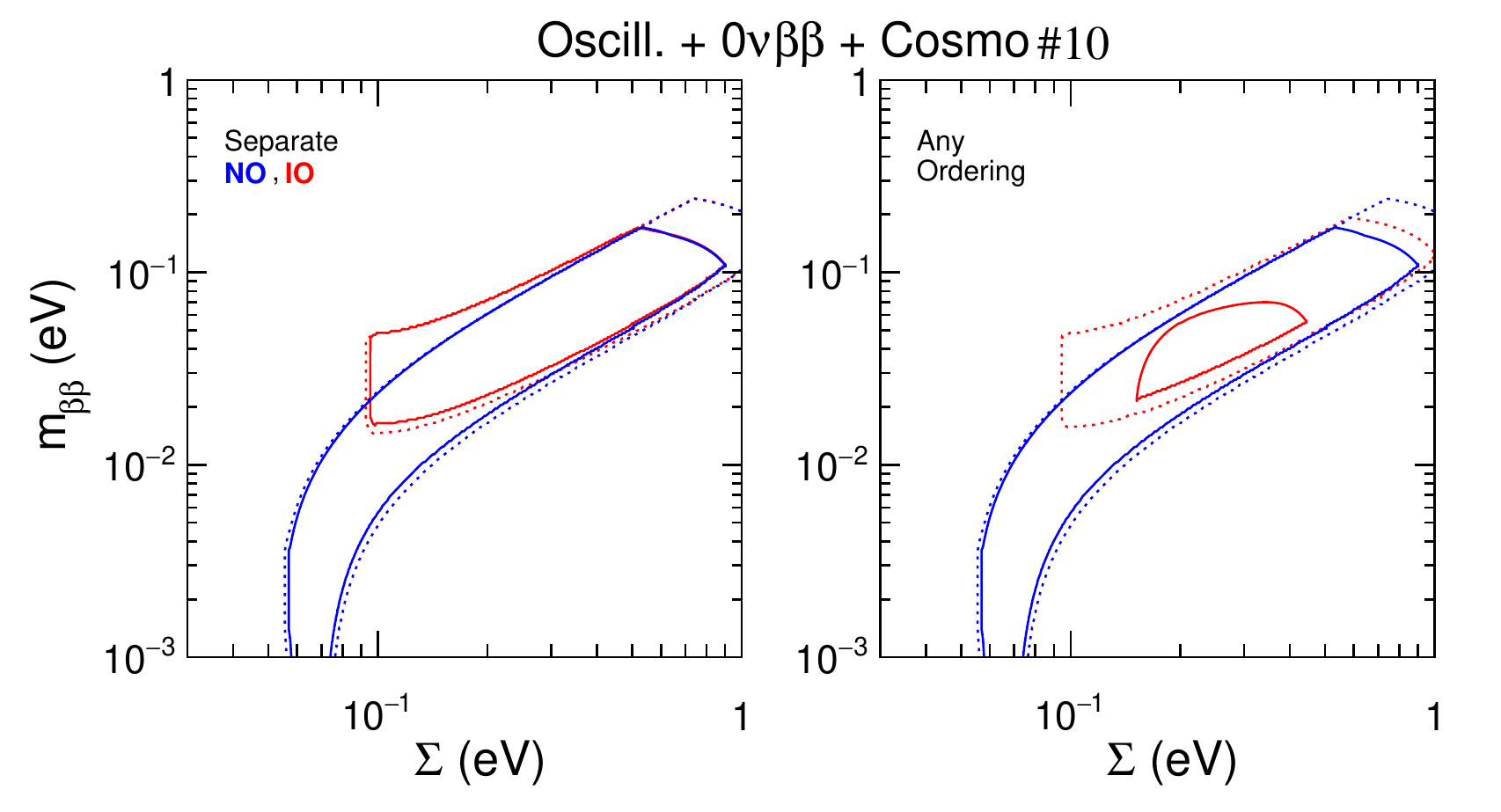}
\caption{\label{fig06}\footnotesize As in Fig.~4, but including the $\chi^2({m_{\beta\beta}})$
function from Fig.~2 and the $\chi^2(\Sigma)$ function from Fig.~3 (for case \#10).} 
\end{minipage}
\end{figure}
%%%%%%%%%%%%%%%%%%%%%%%%%%%%%%%%%%%%%%%%%%%%%%%%%%%%%%%%%%%%%%%%%%%%%%%%%%%%%%%%%%%%%%%%%%

%%%%%%%%%%%%%%%%%%%%%%%%%%%%%%%%%%%%%%%%%%%%%%%%%%%%%%%%%%%%%%%%%%%%%%%%%%%%%%%%%%%%%%%%%%
\begin{figure}[t]
\begin{minipage}[c]{0.96\textwidth}
\includegraphics[width=0.70\textwidth]{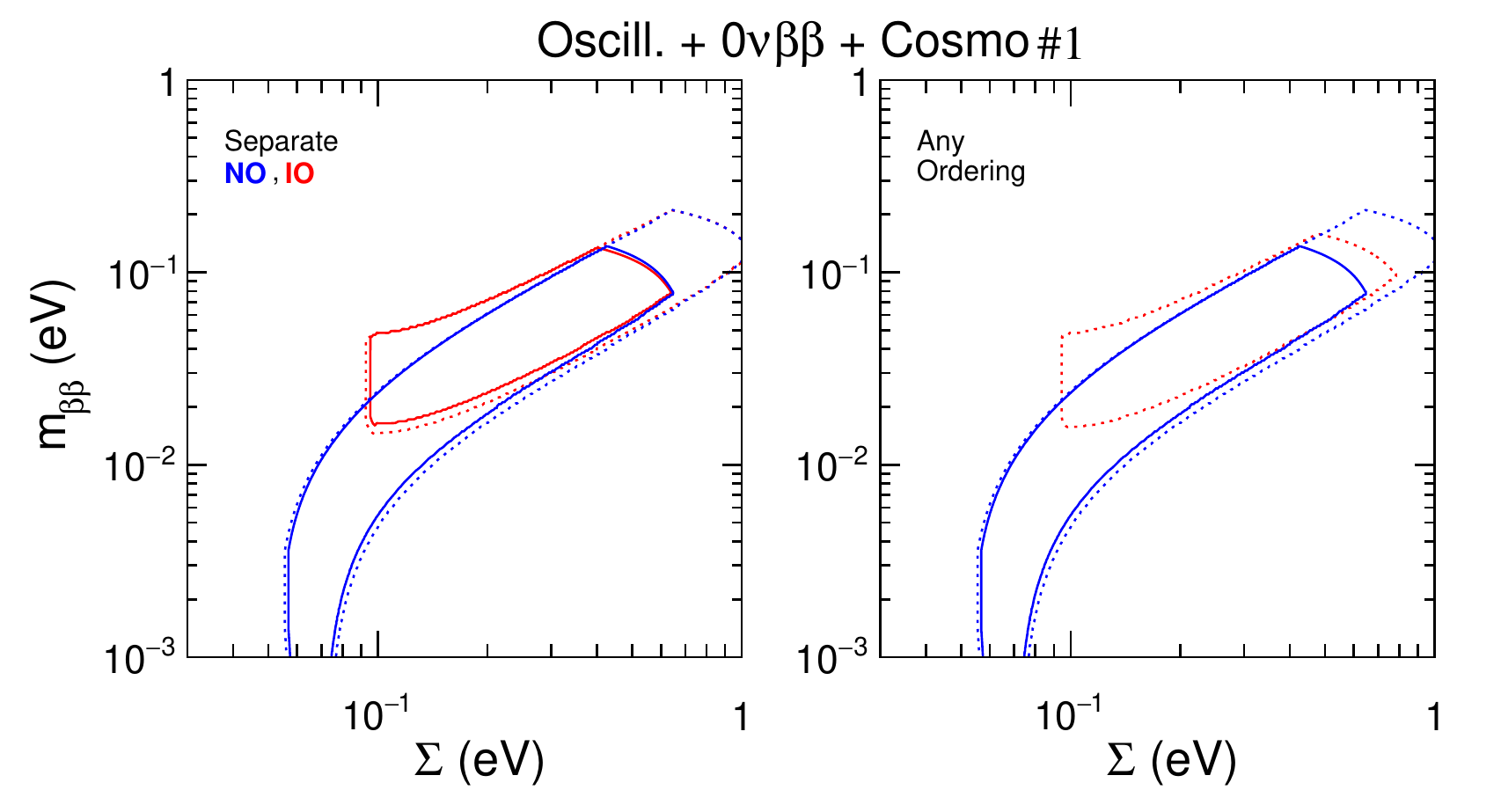}
\caption{\label{fig07}\footnotesize As in Fig.~4, but including the $\chi^2({m_{\beta\beta}})$
function from Fig.~2 and the $\chi^2(\Sigma)$ function from Fig.~3 (for case \#1).} 
\end{minipage}
\end{figure}
%%%%%%%%%%%%%%%%%%%%%%%%%%%%%%%%%%%%%%%%%%%%%%%%%%%%%%%%%%%%%%%%%%%%%%%%%%%%%%%%%%%%%%%%%%

Figure~6 includes, besides oscillation and $0\nu\beta\beta$ constraints,
also the cosmological bounds for the case \#10  in Sec.~II~C (see Table~II). 
The left panel shows a synergic effect of $0\nu\beta\beta$ and cosmological data
in setting a joint $2\sigma$ bound  on $\Sigma$ at the level of $0.9$~eV, to be compared
with the $0\nu\beta\beta$ bound (from Fig.~5) and the cosmological bound (from 
Table~II, case \#10), which are both above 1~eV. A more subtle synergy emerges
from the fact that, for case \#10, the $\chi^2(\Sigma)$ 
function  is minimized at $\sim 0.3$~eV (see Fig.~3), well above the extrema in Eq.~(\ref{lightest2}).
Such a (relatively high) best-fit value for $\Sigma$ implies preferred values 
$m_{\beta\beta}$ around $\mathrm{few}\times 10^{-2}$~eV, as apparent 
for the IO region allowed at $2\sigma$ in the 
right panel. This relatively small IO $2\sigma$ region illustrates qualitatively how the
constraints on ${(\Sigma,\,m_{\beta\beta})}$ would appear 
in the presence of a cosmological measurement (rather than of just upper bounds) for $\Sigma$.

Figures~7, 8 and 9 are analogous to Fig.~6, but refer to the cosmological data sets \#1, \#9, and
\#6 discussed in Sec.~III~C, respectively. Figure~7 shows, once more, the synergy between
comparable $0\nu\beta\beta$ and cosmological bounds on $\Sigma$; indeed, in the left panel,
one reads $\Sigma<0.65$~eV, to be compared with $\Sigma< 0.72$--0.8~eV from
cosmological data only (see Table~II). In the right panel, there is no
IO region allowed at $2\sigma$, since the sum of the $\Delta\chi^2_\mathrm{IO-NO}$ contributions
from oscillation and cosmological data is $3.6 + 0.7 > 4.0$. 

Figure~8 shows, in the left panel,
the transition to a dominance of cosmological constraints on $\Sigma$: the $2\sigma$ bounds
$\Sigma<0.45$-046~eV for case \#9 in Table~II keep $m_{\beta\beta}$ sufficiently small to
suppress any significant impact of current $0\nu\beta\beta$ data in the global fit.  
In the right panel, the relatively deep minimum of $\chi^2(\Sigma)$ evident in Fig.~3
leads to a $2\sigma$ allowed region in IO narrower than in Fig.~6. This region 
shows qualitatively the impact of prospective accurate measurement of $\Sigma$ via
cosmological data.  

\newpage

%%%%%%%%%%%%%%%%%%%%%%%%%%%%%%%%%%%%%%%%%%%%%%%%%%%%%%%%%%%%%%%%%%%%%%%%%%%%%%%%%%%%%%%%%%
\begin{figure}[t]
\begin{minipage}[c]{0.96\textwidth}
\includegraphics[width=0.70\textwidth]{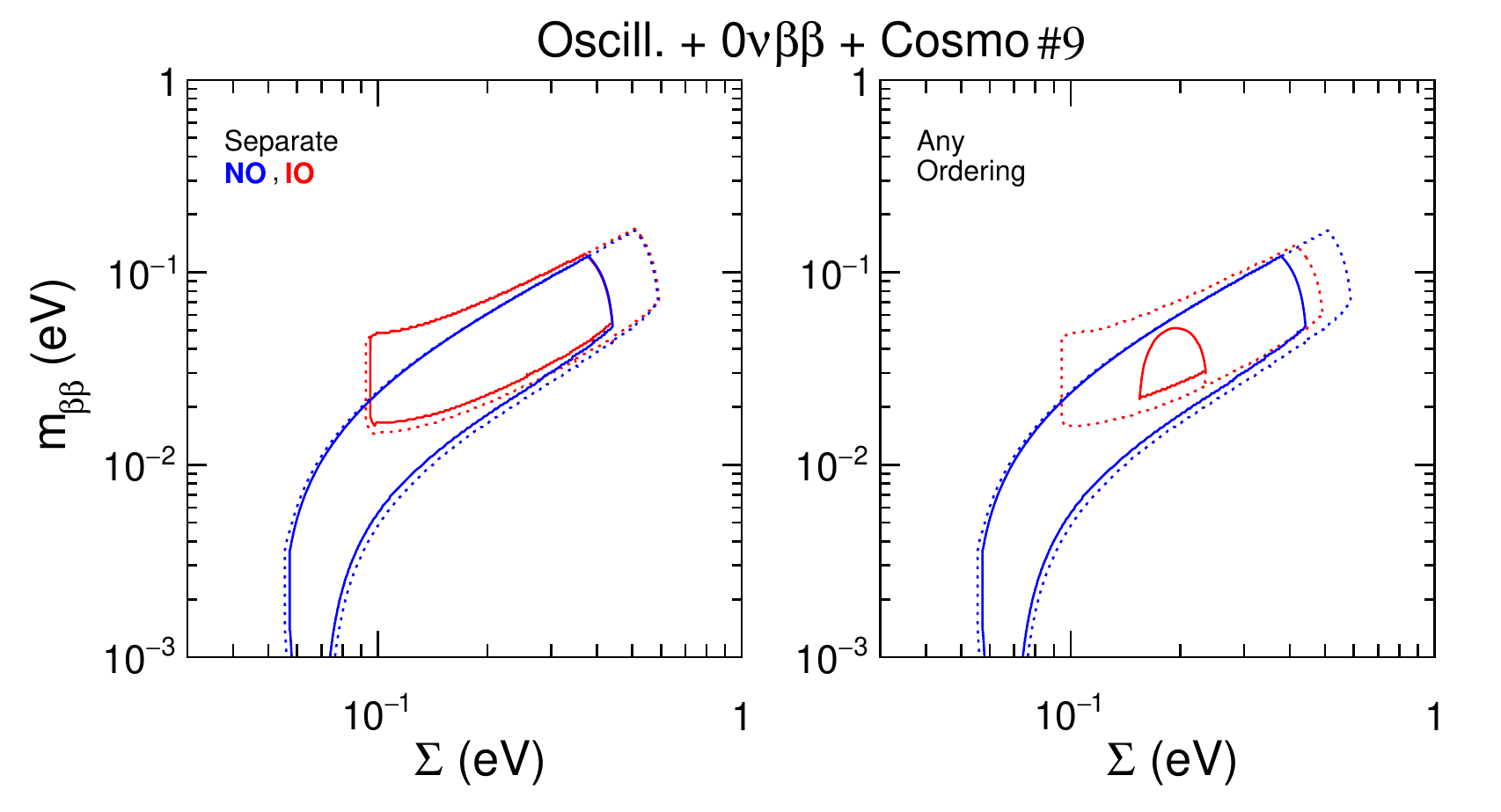}
\caption{\label{fig08}\footnotesize As in Fig.~4, but including the $\chi^2({m_{\beta\beta}})$
function from Fig.~2 and the $\chi^2(\Sigma)$ function from Fig.~3 (for case \#9).} 
\end{minipage}
\end{figure}
%%%%%%%%%%%%%%%%%%%%%%%%%%%%%%%%%%%%%%%%%%%%%%%%%%%%%%%%%%%%%%%%%%%%%%%%%%%%%%%%%%%%%%%%%%

%%%%%%%%%%%%%%%%%%%%%%%%%%%%%%%%%%%%%%%%%%%%%%%%%%%%%%%%%%%%%%%%%%%%%%%%%%%%%%%%%%%%%%%%%%
\begin{figure}[t]
\begin{minipage}[c]{0.96\textwidth}
\includegraphics[width=0.70\textwidth]{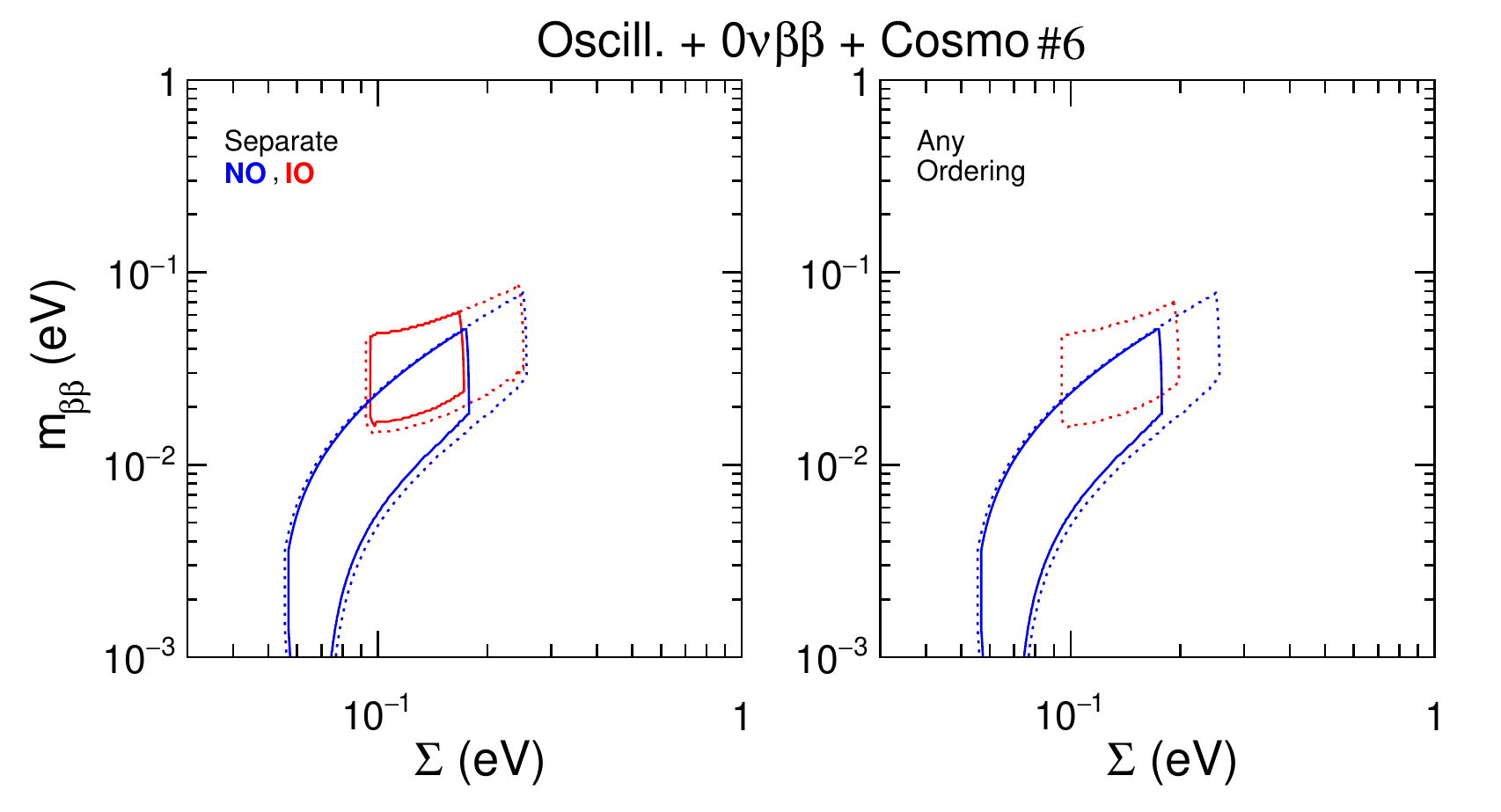}
\caption{\label{fig09}\footnotesize As in Fig.~4, but including the $\chi^2({m_{\beta\beta}})$
function from Fig.~2 and the $\chi^2(\Sigma)$ function from Fig.~3 (for case \#6).} 
\end{minipage}
\end{figure}
%%%%%%%%%%%%%%%%%%%%%%%%%%%%%%%%%%%%%%%%%%%%%%%%%%%%%%%%%%%%%%%%%%%%%%%%%%%%%%%%%%%%%%%%%%

Figure~9 corresponds to the most constraining cosmological case (\#6) in Table~II. In this case, the 
allowed bands are almost vertically cut by the upper bounds on $\Sigma$ from
cosmological data only, with no significant contribution from $0\nu\beta\beta$ constraints. 
Indeed, the allowed values of $m_{\beta\beta}$ are well below the 
$0\nu\beta\beta$ bounds in Eq.~(\ref{mbb-bounds}). Note that,
in the right panel, there is no region allowed at $2\sigma$, since the the global
$\Delta\chi^2_\mathrm{IO-NO}$ exceeds 4 units.

Table~III reports the list of global $\Delta\chi^2_\mathrm{IO-NO}$ values,
numbered according to the cosmological cases in Table~II. 
These values are not always equal to the algebraic sum of the  $\Delta\chi^2$ contributions 
from oscillation data in Eq.~(\ref{preferNO}) and cosmological data in 
Table~II, since the  best-fit
points in the plane $(\Sigma,\,m_{\beta\beta}$)
may be slightly readjusted in NO and IO in the global combination, leading to a small extra
variation ($\delta\chi^2\lesssim 0.4$). This
minor effect in the combination of  $0\nu\beta\beta$ and cosmological  data  
is statistically insignificant at present, but might become more relevant with
future data. In any case, Table~III confirms that 
an overall preference for NO over IO emerges from
the combination of oscillation and nonoscillation data, at the level of $1.9$--$2.1\sigma$.
This is one of the main results of our work.  
\vspace*{-0.7cm}

%===========================================================================
\begin{table}[b]
%\centering
%\captionsetup{width=.96\textwidth}
\caption{\label{Global} \footnotesize Values of $\Delta \chi^2_\mathrm{IO-NO}$  from the global analysis 
of oscillation and non oscillation data (numbered according to the adopted 
cosmological datasets as in Table~II), to be compared 
with the value 3.6 from oscillation data only [Eq.~(\protect\ref{preferNO})].
An overall preference emerges for
NO, at the level of 1.9--$2.1\sigma$.}
\vspace*{0mm}
\centering
\begin{ruledtabular}
\begin{tabular}{rcccccccccccc}
\# & 1 & 2 & 3 & 4 & 5 & 6 & 7 & 8 & 9 & 10 & 11 & 12 \\
\hline
$\Delta\chi^2_\mathrm{IO-NO}$ & 4.3 & 3.8 & 4.4 & 4.2 & 3.9 & 4.4 & 3.6 & 3.7 & 3.8 & 3.7 & 3.8 & 3.9\\
\end{tabular}
\end{ruledtabular}
%\vspace*{.2cm}
\end{table}
%============================================================================
\newpage

We conclude this Section with  a remark on $m_{\beta\beta}$.  In the above figures, 
the $2\sigma$ upper bounds on $m_{\beta\beta}$ decrease from $<0.18$~eV in Fig.~6 (dominated by KamLAND-Zen)
to $<0.06$~eV in Fig.~9 (dominated by cosmology). There are good prospects to further probe this region ---and possibly 
go below it--- with upcoming or planned $0\nu\beta\beta$ experiments \cite{Schw13,Crem14,Bara17}. 
However, unlike $\Sigma$, there is no finite lower bound 
on $m_{\beta\beta}$, since the null value cannot be excluded a priori for unfavorable Majorana phases (see \cite{Xing} and refs.\ therein). 
Conversely, a signal of $m_{\beta\beta}>0$, if accurate enough, might provide some hints or even constraints on such phases 
(see, e.g., \cite{Quiroga}). The identification of Majorana phases as a new source of leptonic CP violation (besides 
the Dirac phase $\delta$) would open new perspectives on the role of leptons in the early universe (see \cite{Riotto} and refs.\ therein).

%%%%%%%%%%%%%%%%%%%%%%%%%%%%%%%%%%%%%%%%%%%%%%%%%%%%%%%%%%%%%%%%%%%%%%%%%%%%%%%%%%%%%%%%%%
\begin{figure}[t]
\begin{minipage}[c]{0.96\textwidth}
\includegraphics[width=0.64\textwidth]{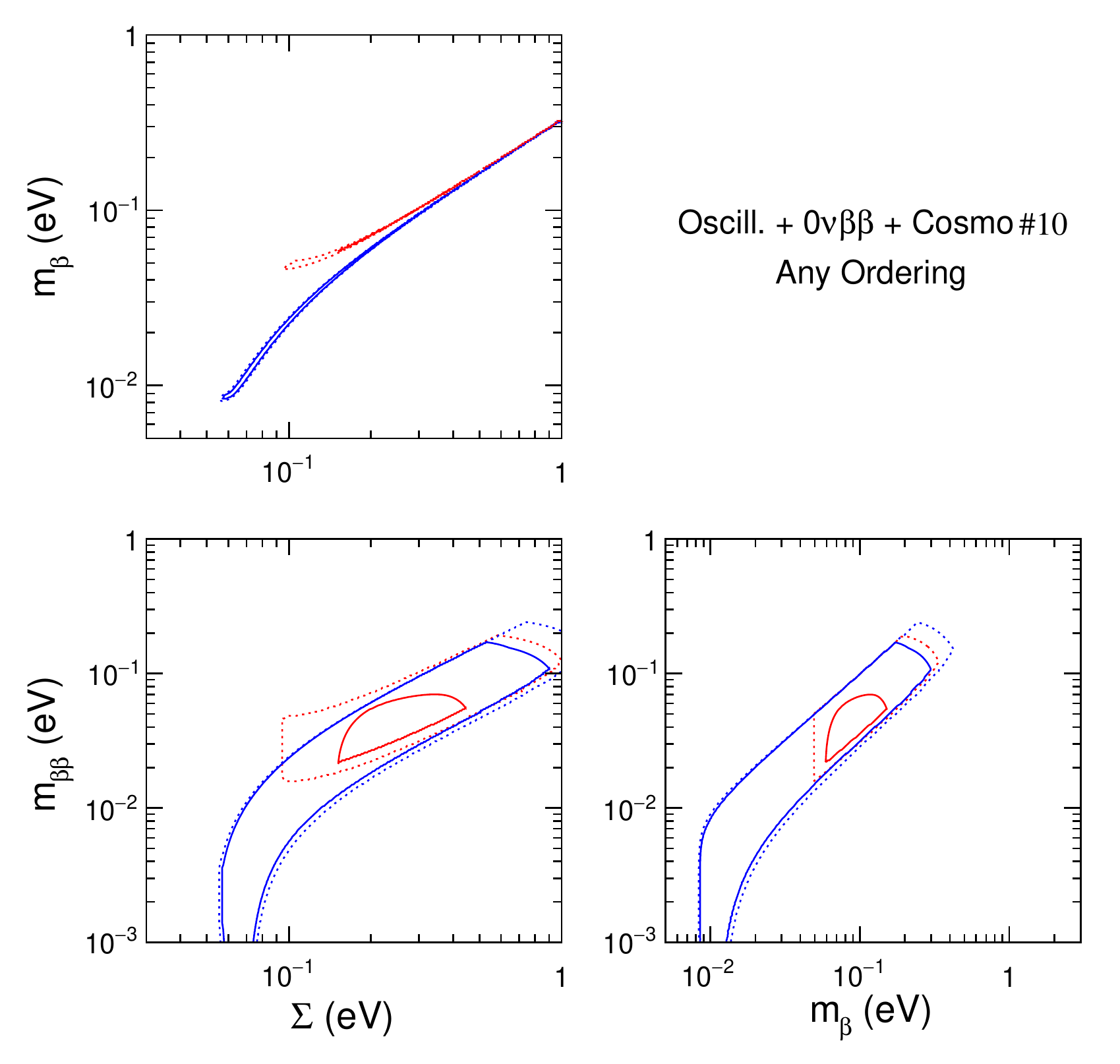}
\caption{\label{fig10}\footnotesize Global $3\nu$ analysis of 
neutrino oscillation and non oscillation data, including the cosmological data set \#10. Bounds  in any mass ordering are shown at $2\sigma$ (solid) and $3\sigma$ (dotted),   
in the planes charted by
any two among the three
absolute mass observables $(\Sigma,\,m_{\beta\beta},\,m_\beta)$.
} 
\end{minipage}
\end{figure}
%%%%%%%%%%%%%%%%%%%%%%%%%%%%%%%%%%%%%%%%%%%%%%%%%%%%%%%%%%%%%%%%%%%%%%%%%%%%%%%%%%%%%%%%%%

\section{Implications for $\bm{m_{\beta}}$}

The results obtained in the previous Section have implications for the discovery potential 
of $\beta$-decay searches, such as the experiment KATRIN \cite{Katrin1,Katrin2,Katrin3}, 
designed to probe the range $m_\beta \gtrsim 0.2$~eV, or future projects, 
envisaged to reach potential sensitivities at or below 0.1~eV \cite{NuMass,Gastaldo,Project8}.
For the sake of brevity, we consider only the case of global fit in any ordering and
for two representative cosmological data sets, namely, \#10 and \#6 in Table~II, that lead to
conservative and aggressive bounds on $\Sigma$, respectively.

Figure~10 shows the bounds on
any two among the three
absolute mass observables $(\Sigma,\,m_{\beta\beta},\,m_\beta)$ for case \#10. 
The $(\Sigma,\,m_{\beta\beta})$ plane is identical to the right
panel of Fig.~6, while the other two planes contain also
the projected bounds on $m_\beta$. The allowed values of $m_\beta$ extend up to
$\sim 0.3$~eV ($2\sigma$) and $\sim 0.4$~eV ($3\sigma$), in the range testable 
by KATRIN; however, a large fraction of the $m_\beta$ allowed 
range, including the preferred IO region at $2\sigma$, 
is below the $0.2$~eV sensitivity goal of this experiment.

%%%%%%%%%%%%%%%%%%%%%%%%%%%%%%%%%%%%%%%%%%%%%%%%%%%%%%%%%%%%%%%%%%%%%%%%%%%%%%%%%%%%%%%%%%
\begin{figure}[t]
\begin{minipage}[c]{0.96\textwidth}
\includegraphics[width=0.64\textwidth]{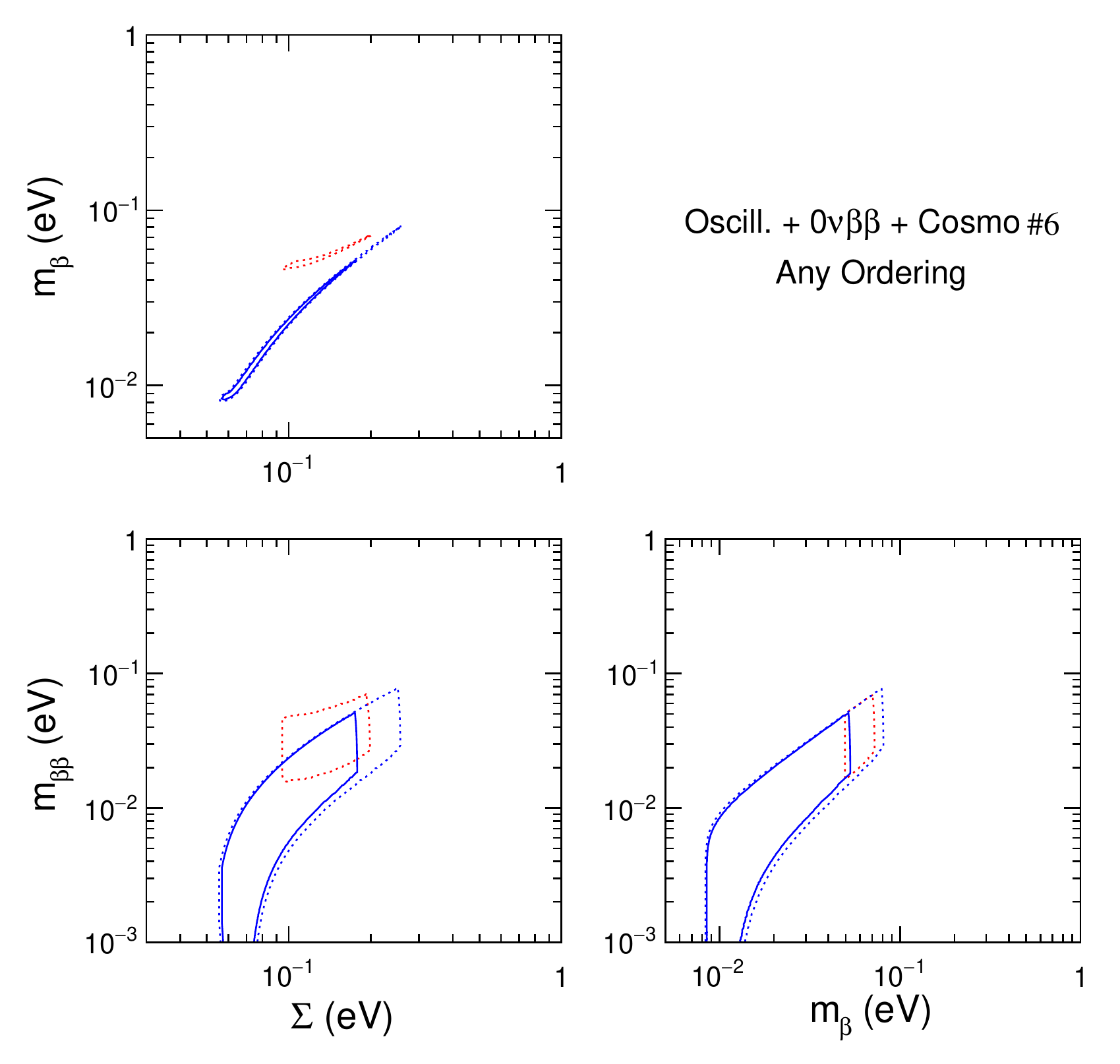}
\caption{\label{fig11}\footnotesize As in Fig.~10, but including 
the cosmological data set \#6.
} 
\end{minipage}
\end{figure}
%%%%%%%%%%%%%%%%%%%%%%%%%%%%%%%%%%%%%%%%%%%%%%%%%%%%%%%%%%%%%%%%%%%%%%%%%%%%%%%%%%%%%%%%%%

Fig.~11 is analogous to Fig.~10, but refers to the cosmological data set \#6.
In this case, the upper bound on $\Sigma$ is very strong, and so is the bound 
on $m_\beta$. Indeed, in the $(\Sigma,\,m_\beta)$ plane, the two allowed branches 
for NO and IO are completely disconnected and could, in principle, be conclusively
discriminated via precise
measurements of $\Sigma$ and $m_\beta$. Unfortunately, the values of $m_\beta$
required by such test are entirely below the KATRIN sensitivity \cite{Katrin1} but, in the long term, they
could be partly probed by planned or envisaged  experimental projects 
\cite{NuMass,Gastaldo,Project8}.

\section{Conclusions}

We have performed a global analysis of oscillation and nonoscillation data within the standard three-neutrino framework.
with particular attention to absolute neutrino masses and their ordering (either normal, NO, or inverted, IO).
Oscillation data have been updated with the latest results, as available at the beginning of 2017.  
$0\nu\beta\beta$ decay bounds have been derived by using recent results from the KamLAND-Zen experiment, together with
a conservative evaluation of nuclear matrix elements and their uncertainties. Cosmological data
from Planck and other experiments have been examined  within the standard $\Lambda$CDM model, with allowance
for nonzero neutrino masses (and eventually for an extra parameter).  The cosmological analysis 
has been performed in a variety of cases, always considering the physical neutrino mass spectra for NO and IO.  

In the global analysis, NO appears to be somewhat favored with 
respect to IO at the level of $1.9$--$2.1\sigma$, mainly by neutrino oscillation data (especially atmospheric), corroborated by cosmological data
in some cases. This intriguing indication, although not statistically mature yet, deserves to be monitored with future data.
Detailed constraints on neutrino mass-mixing parameter have also been obtained via the $\chi^2$ method,
by expanding the parameter space either around separate minima in NO and IO, or around the absolute minimum in any ordering. 
Relevant results have been numerically summarized in Tables~I--III and graphically shown in several figures.
Implications for upcoming oscillation and non-oscillation neutrino experiments, including $\beta$-decay searches, 
have been discussed. 

We emphasize that the above results have been obtained in the standard $3\nu$ framework 
of massive and mixed neutrinos. The experimental search of oscillation phenomena, as well as of signals in the $(m_\beta,\,m_{\beta\beta},\,\Sigma)$
parameter space should, however, be pursued independently of any $3\nu$ expectations, which can
be altered by new (sterile) neutrino states or by new (nonstandard) neutrino interactions,
not considered in this work.

\newpage
%%%%%%%%%%%%%%%%%%%%%%%%%%%%%%%%%%%%%%%%%%%%%%%%%%%%%%%%%%%%%%%%%%%%%%%%%%%%%%%%%%%%%%%%%%%%%%%%%%%%
%%%%%%%%%%%%%%%%%%%%%%%%%%%%%%%%%%%%%%%%%%%%%%%%%%%%%%%%%%%%%%%%%%%%%%%%%%%%%%%%%%%%%%%%%%%%%%%%%%%%

\acknowledgments

This work is supported by the Italian Istituto Nazionale di Fisica 
Nucleare (INFN) and Ministero dell'Istruzione, Universit\`a e Ricerca (MIUR) through the ``Theoretical Astroparticle Physics''  projects. This work has also been performed within the Labex ILP (reference ANR-10-LABX-63) part of the Idex SUPER, and received financial state aid managed by the Agence Nationale de la Recherche, as part of the programme Investissements d'Avenir under the reference ANR-11-IDEX-0004-02.
A.P.\ is supported by the project  {\em Beyond three neutrino families} within the FutureInResearch program,
Fondo di Sviluppo e Coesione 2007-2013, APQ Ricerca Regione Puglia ``Programma regionale a sostegno della specializzazione intelligente e della sostenibilit\`a sociale ed ambientale''. F.C.\ is supported by NSF Grant PHY-1404311 to J.F.~Beacom.
E.L. thanks K.~Inoue and Y.~Koshio for communications about the recent KamLAND-Zen and Super-Kamiokande results, respectively.

%%%%%%%%%%%%%%%%%%%%%%%%%%%%%%%%%%%%%%%%%%%%%%%%%%%%%%%%%%%%%%%%%%%%%%%%%%%%%%%%%%%%%%%%%%%%%%%%


\begin{thebibliography}{99}

\bibitem{PDG2016} 
  C.~Patrignani {\it et al.} [Particle Data Group],
  ``Review of Particle Physics,''
  Chin.\ Phys.\ C {\bf 40}, no. 10, 100001 (2016).
  See the review by K.~Nakamura and S.T.~Petcov on {\em ``Neutrino mass, mixing, and oscillations''\/} therein.
  
\bibitem{Glob06} 
  G.~L.~Fogli, E.~Lisi, A.~Marrone and A.~Palazzo,
  ``Global analysis of three-flavor neutrino masses and mixings,''
  Prog.\ Part.\ Nucl.\ Phys.\  {\bf 57}, 742 (2006)
%  doi:10.1016/j.ppnp.2005.08.002
  [hep-ph/0506083].


\bibitem{Patt15} 
  R.~B.~Patterson,
  ``Prospects for Measurement of the Neutrino Mass Hierarchy,''
  Ann.\ Rev.\ Nucl.\ Part.\ Sci.\  {\bf 65}, 177 (2015)
%  doi:10.1146/annurev-nucl-102014-021916
  [arXiv:1506.07917 [hep-ex]].
  
\bibitem{Qian15} 
  X.~Qian and P.~Vogel,
  ``Neutrino Mass Hierarchy,''
  Prog.\ Part.\ Nucl.\ Phys.\  {\bf 83}, 1 (2015)
%  doi:10.1016/j.ppnp.2015.05.002
  [arXiv:1505.01891 [hep-ex]].



\bibitem{Moha06} 
  R.~N.~Mohapatra and A.~Y.~Smirnov,
  ``Neutrino Mass and New Physics,''
  Ann.\ Rev.\ Nucl.\ Part.\ Sci.\  {\bf 56}, 569 (2006)
%  doi:10.1146/annurev.nucl.56.080805.140534
  [hep-ph/0603118].

\bibitem{Feru15} 
  F.~Feruglio,
  ``Pieces of the Flavour Puzzle,''
  Eur.\ Phys.\ J.\ C {\bf 75}, no. 8, 373 (2015)
%  doi:10.1140/epjc/s10052-015-3576-5
  [arXiv:1503.04071 [hep-ph]].

\bibitem{King15} 
  S.~F.~King,
  ``Models of Neutrino Mass, Mixing and CP Violation,''
  J.\ Phys.\ G {\bf 42}, 123001 (2015)
%  doi:10.1088/0954-3899/42/12/123001
  [arXiv:1510.02091 [hep-ph]].

\bibitem{Octa96} 
  G.~L.~Fogli and E.~Lisi,
  ``Tests of three flavor mixing in long baseline neutrino oscillation experiments,''
  Phys.\ Rev.\ D {\bf 54}, 3667 (1996)
%  doi:10.1103/PhysRevD.54.3667
  [hep-ph/9604415].
 
\bibitem{Capo16} 
  F.~Capozzi, E.~Lisi, A.~Marrone, D.~Montanino and A.~Palazzo,
  ``Neutrino masses and mixings: Status of known and unknown $3\nu$ parameters,''
  Nucl.\ Phys.\ B {\bf 908}, 218 (2016)
%  doi:10.1016/j.nuclphysb.2016.02.016
  [arXiv:1601.07777 [hep-ph]].


 
\bibitem{Marr16} 
  F.~Capozzi, E.~Lisi, A.~Marrone, D.~Montanino and A.~Palazzo,
  ``Status and prospects of global analyses of neutrino mass-mixing parameters,''
  talk presented by A.~Marrone at {\em Neutrino 2016,\/}
 XXVII International Conference on Neutrino Physics and Astrophysics (London, UK, 2016).
 To appear in the Proceedings [Journal of Physics Conference Series (JPCS), 2017].
  Conference website: neutrino2016.iopconfs.org

\bibitem{Conc17} 
  I.~Esteban, M.~C.~Gonzalez-Garcia, M.~Maltoni, I.~Martinez-Soler and T.~Schwetz,
  ``Updated fit to three neutrino mixing: exploring the accelerator-reactor complementarity,''
  JHEP {\bf 1701}, 087 (2017)
%  doi:10.1007/JHEP01(2017)087
  [arXiv:1611.01514 [hep-ph]].

\bibitem{Vall14} 
  D.~V.~Forero, M.~Tortola and J.~W.~F.~Valle,
  ``Neutrino oscillations refitted,''
  Phys.\ Rev.\ D {\bf 90}, no. 9, 093006 (2014)
%  doi:10.1103/PhysRevD.90.093006
  [arXiv:1405.7540 [hep-ph]].


\bibitem{Bile02} 
  S.~M.~Bilenky, C.~Giunti, J.~A.~Grifols and E.~Masso,
  ``Absolute values of neutrino masses: Status and prospects,''
  Phys.\ Rept.\  {\bf 379}, 69 (2003)
%  doi:10.1016/S0370-1573(03)00102-9
  [hep-ph/0211462].

\bibitem{Otte08} 
  E.~W.~Otten and C.~Weinheimer,
  ``Neutrino mass limit from tritium beta decay,''
  Rept.\ Prog.\ Phys.\  {\bf 71}, 086201 (2008)
%  doi:10.1088/0034-4885/71/8/086201
  [arXiv:0909.2104 [hep-ex]].

\bibitem{Viss16}
  S.~Dell'Oro, S.~Marcocci, M.~Viel and F.~Vissani,
  ``Neutrinoless double beta decay: 2015 review,''
  Adv.\ High Energy Phys.\  {\bf 2016}, 2162659 (2016)
%  doi:10.1155/2016/2162659
  [arXiv:1601.07512 [hep-ph]].

\bibitem{Hann10} 
  S.~Hannestad,
  ``Neutrino physics from precision cosmology,''
  Prog.\ Part.\ Nucl.\ Phys.\  {\bf 65}, 185 (2010)
%  doi:10.1016/j.ppnp.2010.07.001
  [arXiv:1007.0658 [hep-ph]].

\bibitem{KamZ16} 
  A.~Gando {\it et al.} [KamLAND-Zen Collaboration],
  ``Search for Majorana Neutrinos near the Inverted Mass Hierarchy Region with KamLAND-Zen,''
  Phys.\ Rev.\ Lett.\  {\bf 117}, no. 8, 082503 (2016);
  Addendum: [Phys.\ Rev.\ Lett.\  {\bf 117}, no. 10, 109903 (2016)]
%  doi:10.1103/PhysRevLett.117.109903, 10.1103/PhysRevLett.117.082503
  [arXiv:1605.02889 [hep-ex]].


\bibitem{Lesg14} 
  J.~Lesgourgues and S.~Pastor,
  ``Neutrino cosmology and Planck,''
  New J.\ Phys.\  {\bf 16}, 065002 (2014)
%  doi:10.1088/1367-2630/16/6/065002
  [arXiv:1404.1740 [hep-ph]].

\bibitem{planckparams2015}
  P.~A.~R.~Ade {\it et al.}  [Planck Collaboration],
  ``Planck 2015 results. XIII. Cosmological parameters,''
  Astron.\ Astrophys.\  {\bf 594}, A13 (2016)
  [arXiv:1502.01589 [astro-ph.CO]].
  %%CITATION = ARXIV:1502.01589;%%
  %302 citations counted in INSPIRE as of 04 juin 2015


\bibitem{NuMass} 
  G.~J.~Barker {\it et al.},
  ``The Future of Neutrino Mass Measurements: Terrestrial, Astrophysical, and Cosmological Measurements in the Next Decade. Highlights of the {\em NuMass 2013\/} Workshop (Milano, Italy, February 4--7, 2013),''
  arXiv:1309.7810 [hep-ex].
  %%CITATION = ARXIV:1309.7810;%%
  %5 citations counted in INSPIRE as of 28 Feb 2017




\bibitem{Melc04} 
  G.~L.~Fogli, E.~Lisi, A.~Marrone, A.~Melchiorri, A.~Palazzo, P.~Serra and J.~Silk,
  ``Observables sensitive to absolute neutrino masses: Constraints and correlations from world neutrino data,''
  Phys.\ Rev.\ D {\bf 70}, 113003 (2004)
%  doi:10.1103/PhysRevD.70.113003
  [hep-ph/0408045].

\bibitem{Melc06} 
  G.~L.~Fogli, E.~Lisi, A.~Marrone, A.~Melchiorri, A.~Palazzo, P.~Serra, J.~Silk and A.~Slosar,
  ``Observables sensitive to absolute neutrino masses: A Reappraisal after WMAP-3y and first MINOS results,''
  Phys.\ Rev.\ D {\bf 75}, 053001 (2007)
%  doi:10.1103/PhysRevD.75.053001
  [hep-ph/0608060].

\bibitem{Melc08}
  G.~L.~Fogli {\it et al.},
  ``Observables sensitive to absolute neutrino masses.~II,''
  Phys.\ Rev.\ D {\bf 78}, 033010 (2008)
%  doi:10.1103/PhysRevD.78.033010
  [arXiv:0805.2517 [hep-ph]].

\bibitem{Stat16}   G.~Cowan, Review on {\em ``Statistics''\/} in \protect\cite{PDG2016}. See also
F.~James and M.~Roos,
  ``Minuit: A System for Function Minimization and Analysis of the Parameter Errors and Correlations,''
  Comput.\ Phys.\ Commun.\  {\bf 10}, 343 (1975).
%  doi:10.1016/0010-4655(75)90039-9
 
\bibitem{Blen13} 
  M.~Blennow, P.~Coloma, P.~Huber and T.~Schwetz,
  ``Quantifying the sensitivity of oscillation experiments to the neutrino mass ordering,''
  JHEP {\bf 1403}, 028 (2014)
%  doi:10.1007/JHEP03(2014)028
  [arXiv:1311.1822 [hep-ph]].

\bibitem{Bahc98} 
  J.~N.~Bahcall, P.~I.~Krastev and A.~Y.~Smirnov,
  ``Where do we stand with solar neutrino oscillations?,''
  Phys.\ Rev.\ D {\bf 58}, 096016 (1998)
%  doi:10.1103/PhysRevD.58.096016
  [hep-ph/9807216].

\bibitem{Petc98} 
  S.~T.~Petcov,
  ``The Solar neutrino problem and solar neutrino oscillations in vacuum and in matter,''
  Lect.\ Notes Phys.\  {\bf 512}, 281 (1998)
%  doi:10.1007/BFb0106897
  [hep-ph/9806466].
  
  
\bibitem{Mont96} 
  G.~L.~Fogli, E.~Lisi and D.~Montanino,
  ``Matter enhanced three flavor oscillations and the solar neutrino problem,''
  Phys.\ Rev.\ D {\bf 54}, 2048 (1996)
%  doi:10.1103/PhysRevD.54.2048
  [hep-ph/9605273].

  
\bibitem{Mura00} 
  A.~de Gouvea, A.~Friedland and H.~Murayama,
  ``The Dark side of the solar neutrino parameter space,''
  Phys.\ Lett.\ B {\bf 490}, 125 (2000)
%  doi:10.1016/S0370-2693(00)00989-8
  [hep-ph/0002064].

\bibitem{Frie00} 
  A.~Friedland,
  ``MSW effects in vacuum oscillations,''
  Phys.\ Rev.\ Lett.\  {\bf 85}, 936 (2000)
%  doi:10.1103/PhysRevLett.85.936
  [hep-ph/0002063].

\bibitem{Lisi00} 
  E.~Lisi, A.~Marrone, D.~Montanino, A.~Palazzo and S.~T.~Petcov,
  ``Analytical description of quasivacuum oscillations of solar neutrinos,''
  Phys.\ Rev.\ D {\bf 63}, 093002 (2001)
%  doi:10.1103/PhysRevD.63.093002
  [hep-ph/0011306].


\bibitem{Pena01} 
  M.~C.~Gonzalez-Garcia, M.~Maltoni, C.~Pena-Garay and J.~W.~F.~Valle,
  ``Global three neutrino oscillation analysis of neutrino data,''
  Phys.\ Rev.\ D {\bf 63}, 033005 (2001)
%  doi:10.1103/PhysRevD.63.033005
  [hep-ph/0009350].


\bibitem{Gett02} 
  G.~L.~Fogli, E.~Lisi, A.~Marrone, D.~Montanino and A.~Palazzo,
  ``Getting the most from the statistical analysis of solar neutrino oscillations,''
  Phys.\ Rev.\ D {\bf 66}, 053010 (2002)
%  doi:10.1103/PhysRevD.66.053010
  [hep-ph/0206162].

\bibitem{Bahc04} 
  J.~N.~Bahcall,
  ``Solving the mystery of the missing neutrinos,''
  physics/0406040. Also available at the website
  www.nobelprize.org/nobel\_prizes/themes/physics/bahcall

\bibitem{Lyons}
L.\ Lyons, ``Selecting between two hypotheses,'' Oxford University preprint OUNP-99-12 (1999);
 ``Statistical techniques in high energy physics,'' in {\em ACAT 2000,\/} VII International Workshop
on Advanced Computing and Analysis Techniques in Physics Research (Batavia, IL, 2000),
AIP Conference Proceedings 583, 31 (2001).

\bibitem{Bile17} 
  S.~M.~Bilenky, F.~Capozzi and S.~T.~Petcov,
  ``An Alternative Method of Determining the Neutrino Mass Ordering in Reactor Neutrino Experiments,''
  arXiv:1701.06328 [hep-ph].

\bibitem{Juno14} 
  F.~Capozzi, E.~Lisi and A.~Marrone,
  ``Neutrino mass hierarchy and electron neutrino oscillation parameters with one hundred thousand reactor events,''
  Phys.\ Rev.\ D {\bf 89}, no. 1, 013001 (2014)
%  doi:10.1103/PhysRevD.89.013001
  [arXiv:1309.1638 [hep-ph]].
  
\bibitem{Ciuffoli} 
  E.~Ciuffoli, J.~Evslin and X.~Zhang,
  ``Confidence in a neutrino mass hierarchy determination,''
  JHEP {\bf 1401}, 095 (2014)
%  doi:10.1007/JHEP01(2014)095
  [arXiv:1305.5150 [hep-ph]].


\bibitem{Blennow} 
  M.~Blennow,
  ``On the Bayesian approach to neutrino mass ordering,''
  JHEP {\bf 1401}, 139 (2014)
%  doi:10.1007/JHEP01(2014)139
  [arXiv:1311.3183 [hep-ph]].

\bibitem{Stanco} 
  L.~Stanco, S.~Dusini and M.~Tenti,
  ``Determination of the neutrino mass hierarchy with a new statistical method,''
  arXiv:1606.09454 [hep-ph].


  

\bibitem{NuFit} 
  M.~C.~Gonzalez-Garcia, M.~Maltoni and T.~Schwetz,
  ``Updated fit to three neutrino mixing: status of leptonic CP violation,''
  JHEP {\bf 1411}, 052 (2014)
%  doi:10.1007/JHEP11(2014)052
  [arXiv:1409.5439 [hep-ph]].


\bibitem{T2K2017} 
  K.~Abe {\it et al.} [T2K Collaboration],
  ``First combined analysis of neutrino and antineutrino oscillations at T2K,''
  arXiv:1701.00432 [hep-ex].
  
  
\bibitem{NOvA2016}  
  P.~Adamson {\it et al.} [NOvA Collaboration],
  ``Constraints on oscillation parameters from $\nu_e$ appearance and $\nu_\mu$ disappearance in NOvA,''
  arXiv:1703.03328 [hep-ex]. See also P.~Vahle in {\em Neutrino 2016\/} \protect\cite{Marr16}. 
  %%CITATION = ARXIV:1703.03328;%%
  
\bibitem{NOvA2017} 
  P.~Adamson {\it et al.} [NOvA Collaboration],
  ``Measurement of the neutrino mixing angle $\theta_{23}$ in NOvA,''
  [arXiv:1701.05891 [hep-ex]]. 
  %%CITATION = ARXIV:1701.05891;%%
  
\bibitem{DB2017} 
  F.~P.~An {\it et al.} [Daya Bay Collaboration],
  ``Improved Measurement of the Reactor Antineutrino Flux and Spectrum at Daya Bay,''
  Chin.\ Phys.\ C {\bf 2017}, 41
%  doi:10.1088/1674-1137/41/1/013002
  [arXiv:1607.05378 [hep-ex]];
  ``Measurement of electron antineutrino oscillation based on 1230 days of operation of the Daya Bay experiment,''
  arXiv:1610.04802 [hep-ex].
 
  
\bibitem{SKatm1}
	Z.\ Li, ``Atmospheric Neutrino Results from Super-Kamiokande,'' in {\em ICHEP 2016,\/}
	 38th International Conference on High Energy Physics (Chicago, IL, 2016), PoS(ICHEP2016)461.  
	 
\bibitem{SKatm2}
	Y.\ Koshio, ``Solar and atmospheric neutrino oscillations in Super-Kamiokande,'' in {\em NOW 2016,\/} 
  	International Neutrino Oscillation Workshop (Otranto, Italy, 2016), PoS(NOW2016)001.
    Workshop Website: www.ba.infn.it/now
    
\bibitem{Hint} 
  G.~L.~Fogli, E.~Lisi, A.~Marrone, A.~Palazzo and A.~M.~Rotunno,
  ``Hints of $\theta_{13}>0$ from global neutrino data analysis,''
  Phys.\ Rev.\ Lett.\  {\bf 101}, 141801 (2008)
%  doi:10.1103/PhysRevLett.101.141801
  [arXiv:0806.2649 [hep-ph]].

\bibitem{Kearns}
E.D.\ Kearns,  ``3-flavor oscillations with current and future atmospheric experiments,''
contribution to the April 2017 Meeting of the American Physical Society (Washington, DC, 2017). Website:
meetings.aps.org/Meeting/APR17/Session/J10.3

\bibitem{Giun15}
S.~M.~Bilenky and C.~Giunti,
  ``Neutrinoless Double-Beta Decay: a Probe of Physics Beyond the Standard Model,''
  Int.\ J.\ Mod.\ Phys.\ A {\bf 30}, 1530001 (2015)
%  doi:10.1142/S0217751X1530001X
  [arXiv:1411.4791 [hep-ph]].

\bibitem{Paes15} 
  H.~P{\"a}s and W.~Rodejohann,
  ``Neutrinoless Double Beta Decay,''
  New J.\ Phys.\  {\bf 17}, no. 11, 115010 (2015)
%  doi:10.1088/1367-2630/17/11/115010
  [arXiv:1507.00170 [hep-ph]].    
 
\bibitem{Verg16} 
  J.~D.~Vergados, H.~Ejiri and F.~Simkovic,
  ``Neutrinoless double beta decay and neutrino mass,''
  Int.\ J.\ Mod.\ Phys.\ E {\bf 25}, no. 11, 1630007 (2016)
%  doi:10.1142/S0218301316300071
  [arXiv:1612.02924 [hep-ph]].

\bibitem{Schw13} 
  B.~Schwingenheuer,
  ``Status and prospects of searches for neutrinoless double beta decay,''
  Annalen Phys.\  {\bf 525}, 269 (2013)
%  doi:10.1002/andp.201200222
  [arXiv:1210.7432 [hep-ex]].
 

\bibitem{Crem14} 
  O.~Cremonesi and M.~Pavan,
  ``Challenges in Double Beta Decay,''
  Adv.\ High Energy Phys.\  {\bf 2014}, 951432 (2014)
%  doi:10.1155/2014/951432
  [arXiv:1310.4692 [physics.ins-det]].
   
\bibitem{Bara17}
A.~S.~Barabash,
  ``Brief review of double beta decay experiments,''
  arXiv:1702.06340 [nucl-ex]. Contribution to {\em ICPPA 2016\/},
  2nd International Conference on Particle Physics and Astrophysics 	
  (Moscow, Russia, 2016). 
  

\bibitem{Voge12} 
  P.~Vogel,
  ``Nuclear structure and double beta decay,''
  J.\ Phys.\ G {\bf 39}, 124002 (2012)
%  doi:10.1088/0954-3899/39/12/124002
  [arXiv:1208.1992 [nucl-th]].
 
\bibitem{Suho15} 
  J.~Hyv�rinen and J.~Suhonen,
  ``Nuclear matrix elements for $0\nu\beta\beta$ decays with light or heavy Majorana-neutrino exchange,''
  Phys.\ Rev.\ C {\bf 91}, no. 2, 024613 (2015).
%  doi:10.1103/PhysRevC.91.024613
   
  
\bibitem{Enge17} 
  J.~Engel and J.~Menendez,
  ``Status and Future of Nuclear Matrix Elements for Neutrinoless Double-Beta Decay: A Review,''
  arXiv:1610.06548 [nucl-th].
  
  

\bibitem{KLchi2}
	J.\ Ouellet, ``Results and Status from KamLAND-Zen,'' in {\em ICHEP 2016,\/}
	 38th International Conference on High Energy Physics (Chicago, IL, 2016), PoS(ICHEP2016)492.
	 See also the talk online at indico.cern.ch/event/432527  
	
\bibitem{Rotu15} 
  E.~Lisi, A.~Rotunno and F.~Simkovic,
  ``Degeneracies of particle and nuclear physics uncertainties in neutrinoless $\beta \beta$ decay,''
  Phys.\ Rev.\ D {\bf 92}, no. 9, 093004 (2015)
%  doi:10.1103/PhysRevD.92.093004
  [arXiv:1506.04058 [hep-ph]].
  
\bibitem{Bare15} 
  J.~Barea, J.~Kotila and F.~Iachello,
  ``Nuclear matrix elements for double-$\beta$ decay,''
  Phys.\ Rev.\ C {\bf 87}, no. 1, 014315 (2013)
%  doi:10.1103/PhysRevC.87.014315
  [arXiv:1301.4203 [nucl-th]].
  ``$0\nu\beta\beta$ and $2\nu\beta\beta$ nuclear matrix elements in the interacting boson model with isospin restoration,''
  Phys.\ Rev.\ C {\bf 91}, no. 3, 034304 (2015)
%  doi:10.1103/PhysRevC.91.034304
  [arXiv:1506.08530 [nucl-th]].


\bibitem{QRPA08} 
  A.~Faessler, G.~L.~Fogli, E.~Lisi, V.~Rodin, A.~M.~Rotunno and F.~Simkovic,
  ``QRPA uncertainties and their correlations in the analysis of $0\nu\beta\beta$ decay,''
  Phys.\ Rev.\ D {\bf 79}, 053001 (2009)
%  doi:10.1103/PhysRevD.79.053001
  [arXiv:0810.5733 [hep-ph]].

%%%%%%%%%%%%%%%%%%%%%%%%%%%%%%%%%%%%%%%%%%%%%%%%%%%%%%%%%%%%%%%%%%%%%%%%%%
%%%%%%%%%%%%%%%%%%%%%%%%%%%%%%%%%%%%%%%%%%%%%%%%%%%%%%%%%%%%%%%%%%%%%%%%


%% Theory

\bibitem{Hu:1997mj}
  W.~Hu, D.~J.~Eisenstein and M.~Tegmark,
  ``Weighing neutrinos with galaxy surveys,''
  Phys.\ Rev.\ Lett.\  {\bf 80}, 5255 (1998)
%  doi:10.1103/PhysRevLett.80.5255
  [astro-ph/9712057].
  %%CITATION = doi:10.1103/PhysRevLett.80.5255;%%
  %333 citations counted in INSPIRE as of 08 Mar 2017
  
\bibitem{Dolgov:2002wy}
  A.~D.~Dolgov,
  ``Neutrinos in cosmology,''
  Phys.\ Rept.\  {\bf 370} (2002) 333
%  doi:10.1016/S0370-1573(02)00139-4
  [hep-ph/0202122].
  %%CITATION = doi:10.1016/S0370-1573(02)00139-4;%%
  %402 citations counted in INSPIRE as of 08 Mar 2017

\bibitem{Lesgourgues:2006nd}
  J.~Lesgourgues and S.~Pastor,
  ``Massive neutrinos and cosmology,''
  Phys.\ Rept.\  {\bf 429}, 3017 (2006)
%  doi:10.1016/j.physrep.2006.04.001
  [astro-ph/0603494].
  %%CITATION = doi:10.1016/j.physrep.2006.04.001;%%
  %574 citations counted in INSPIRE as of 08 Mar 2017
  
%  \bibitem{Hannestad:2010kz}
%  S.~Hannestad,
%  %``Neutrino physics from precision cosmology,''
%  Prog.\ Part.\ Nucl.\ Phys.\  {\bf 65} (2010) 185
%  doi:10.1016/j.ppnp.2010.07.001
%  [arXiv:1007.0658 [hep-ph]].
%  %%CITATION = doi:10.1016/j.ppnp.2010.07.001;%%
%  %75 citations counted in INSPIRE as of 08 Mar 2017
  
  \bibitem{Abazajian:2011dt}
  K.~N.~Abazajian {\it et al.},
  ``Cosmological and Astrophysical Neutrino Mass Measurements,''
  Astropart.\ Phys.\  {\bf 35}, 177 (2011)
%  doi:10.1016/j.astropartphys.2011.07.002
  [arXiv:1103.5083 [astro-ph.CO]].
  %%CITATION = doi:10.1016/j.astropartphys.2011.07.002;%%
  %136 citations counted in INSPIRE as of 08 Mar 2017

  \bibitem{Lesgourgues:2012uu}
  J.~Lesgourgues and S.~Pastor,
  ``Neutrino mass from Cosmology,''
  Adv.\ High Energy Phys.\  {\bf 2012}, 608515 (2012) 
%  doi:10.1155/2012/608515
  [arXiv:1212.6154 [hep-ph]].
  %%CITATION = doi:10.1155/2012/608515;%%
  %101 citations counted in INSPIRE as of 08 Mar 2017
  
  
%% Bounds recenti

\bibitem{Vagnozzi:2017ovm} 
  S.~Vagnozzi, E.~Giusarma, O.~Mena, K.~Freese, M.~Gerbino, S.~Ho and M.~Lattanzi,
  ``Unveiling $\nu$ secrets with cosmological data: neutrino masses and mass hierarchy,''
  arXiv:1701.08172 [astro-ph.CO].

\bibitem{Giusarma:2016phn} 
  E.~Giusarma, M.~Gerbino, O.~Mena, S.~Vagnozzi, S.~Ho and K.~Freese,
  ``Improvement of cosmological neutrino mass bounds,''
  Phys.\ Rev.\ D {\bf 94}, no. 8, 083522 (2016)
%  doi:10.1103/PhysRevD.94.083522
  [arXiv:1605.04320 [astro-ph.CO]].

\bibitem{Xu:2016ddc} 
  L.~Xu and Q.~G.~Huang,
  ``Detecting the Neutrinos Mass Hierarchy from Cosmological Data,''
  arXiv:1611.05178 [astro-ph.CO].
  %%CITATION = ARXIV:1611.05178;%%
  %2 citations counted in INSPIRE as of 10 Feb 2017

\bibitem{Huang:2015wrx} 
  Q.~G.~Huang, K.~Wang and S.~Wang,
  ``Constraints on the neutrino mass and mass hierarchy from cosmological observations,''
  Eur.\ Phys.\ J.\ C {\bf 76}, no. 9, 489 (2016)
%  doi:10.1140/epjc/s10052-016-4334-z
  [arXiv:1512.05899 [astro-ph.CO]].
  %%CITATION = doi:10.1140/epjc/s10052-016-4334-z;%%
  %13 citations counted in INSPIRE as of 10 Feb 2017

\bibitem{DiValentino:2015sam} 
  E.~Di Valentino, E.~Giusarma, O.~Mena, A.~Melchiorri and J.~Silk,
  ``Cosmological limits on neutrino unknowns versus low redshift priors,''
  Phys.\ Rev.\ D {\bf 93}, no. 8, 083527 (2016)
%  doi:10.1103/PhysRevD.93.083527
  [arXiv:1511.00975 [astro-ph.CO]].
  %%CITATION = doi:10.1103/PhysRevD.93.083527;%%
  %24 citations counted in INSPIRE as of 10 Feb 2017

\bibitem{Cuesta:2015iho} 
  A.~J.~Cuesta, V.~Niro and L.~Verde,
  ``Neutrino mass limits: robust information from the power spectrum of galaxy surveys,''
  Phys.\ Dark Univ.\  {\bf 13}, 77 (2016)
%  doi:10.1016/j.dark.2016.04.005
  [arXiv:1511.05983 [astro-ph.CO]].
  %%CITATION = doi:10.1016/j.dark.2016.04.005;%%
  %35 citations counted in INSPIRE as of 10 Feb 2017

%forecasts
%\cite{Archidiacono:2016lnv}
\bibitem{Archidiacono:2016lnv} 
  M.~Archidiacono, T.~Brinckmann, J.~Lesgourgues and V.~Poulin,
  ``Physical effects involved in the measurements of neutrino masses with future cosmological data,''
  arXiv:1610.09852 [astro-ph.CO].
  %%CITATION = ARXIV:1610.09852;%%
  %4 citations counted in INSPIRE as of 10 Feb 2017

\bibitem{Oyama:2015gma} 
  Y.~Oyama, K.~Kohri and M.~Hazumi,
  ``Constraints on the neutrino parameters by future cosmological 21 cm line and precise CMB polarization observations,''
  JCAP {\bf 1602}, no. 02, 008 (2016)
%  doi:10.1088/1475-7516/2016/02/008
  [arXiv:1510.03806 [astro-ph.CO]].
  %%CITATION = doi:10.1088/1475-7516/2016/02/008;%%
  %11 citations counted in INSPIRE as of 10 Feb 2017

% forecasts
%\cite{Allison:2015qca}
\bibitem{Allison:2015qca} 
  R.~Allison, P.~Caucal, E.~Calabrese, J.~Dunkley and T.~Louis,
  ``Towards a cosmological neutrino mass detection,''
  Phys.\ Rev.\ D {\bf 92}, no. 12, 123535 (2015)
%  doi:10.1103/PhysRevD.92.123535
  [arXiv:1509.07471 [astro-ph.CO]].
  %%CITATION = doi:10.1103/PhysRevD.92.123535;%%
  %29 citations counted in INSPIRE as of 10 Feb 2017

\bibitem{DiValentino:2016foa}
  E.~Di Valentino {\it et al.} [CORE Collaboration],
  ``Exploring Cosmic Origins with CORE: Cosmological Parameters,''
  arXiv:1612.00021 [astro-ph.CO].
  %%CITATION = ARXIV:1612.00021;%%
  %13 citations counted in INSPIRE as of 08 Mar 2017

% forecasts
%\cite{Banerjee:2016suz}
\bibitem{Banerjee:2016suz} 
  A.~Banerjee, B.~Jain, N.~Dalal and J.~Shelton,
  ``Tests of Neutrino and Dark Radiation Models from Galaxy and CMB surveys,''
  arXiv:1612.07126 [astro-ph.CO].
  %%CITATION = ARXIV:1612.07126;%%
  %1 citations counted in INSPIRE as of 10 Feb 2017

%\cite{Hamann:2012fe}
\bibitem{Hamann:2012fe} 
  J.~Hamann, S.~Hannestad and Y.~Y.~Y.~Wong,
  ``Measuring neutrino masses with a future galaxy survey,''
  JCAP {\bf 1211}, 052 (2012)
%  doi:10.1088/1475-7516/2012/11/052
  [arXiv:1209.1043 [astro-ph.CO]].
  %%CITATION = doi:10.1088/1475-7516/2012/11/052;%%
  %53 citations counted in INSPIRE as of 10 Feb 2017

%dependence of the neutrino mass bounds on the assumed model


\bibitem{DiValentino:2015ola}
  E.~Di Valentino, A.~Melchiorri and J.~Silk,
  ``Beyond six parameters: extending $\Lambda$CDM,''
  Phys.\ Rev.\ D {\bf 92}  n.~12,  121302 (2015)
 % doi:10.1103/PhysRevD.92.121302
  [arXiv:1507.06646 [astro-ph.CO]].
  %%CITATION = doi:10.1103/PhysRevD.92.121302;%%
  %14 citations counted in INSPIRE as of 08 Mar 2017

\bibitem{DiValentino:2016hlg}
  E.~Di Valentino, A.~Melchiorri and J.~Silk,
  ``Reconciling Planck with the local value of $H_0$ in extended parameter space,''
  Phys.\ Lett.\ B {\bf 761}, 242 (2016)
%  doi:10.1016/j.physletb.2016.08.043
  [arXiv:1606.00634 [astro-ph.CO]].
  %%CITATION = doi:10.1016/j.physletb.2016.08.043;%%
  %30 citations counted in INSPIRE as of 08 Mar 2017
  
\bibitem{Motohashi:2012wc}
  H.~Motohashi, A.~A.~Starobinsky and J.~Yokoyama,
  ``Cosmology Based on f(R) Gravity Admits 1 eV Sterile Neutrinos,''
  Phys.\ Rev.\ Lett.\  {\bf 110}  n.~12,  121302 (2013)
%  doi:10.1103/PhysRevLett.110.121302
  [arXiv:1203.6828 [astro-ph.CO]].
  %%CITATION = doi:10.1103/PhysRevLett.110.121302;%%
  %35 citations counted in INSPIRE as of 08 Mar 2017

\bibitem{Hu:2014sea}
  B.~Hu, M.~Raveri, A.~Silvestri and N.~Frusciante,
  ``Exploring massive neutrinos in dark cosmologies with EFTCAMB/EFTCosmoMC,''
  Phys.\ Rev.\ D {\bf 91} no.6,  063524 (2015) 
%  doi:10.1103/PhysRevD.91.063524
  [arXiv:1410.5807 [astro-ph.CO]].
  %%CITATION = doi:10.1103/PhysRevD.91.063524;%%
  %17 citations counted in INSPIRE as of 08 Mar 2017

\bibitem{Wang:2016tsz}
  S.~Wang, Y.~F.~Wang, D.~M.~Xia and X.~Zhang,
  ``Impacts of dark energy on weighing neutrinos: mass hierarchies considered,''
  Phys.\ Rev.\ D {\bf 94}  n.~8,  083519 (2016)
%  doi:10.1103/PhysRevD.94.083519
  [arXiv:1608.00672 [astro-ph.CO]].
  %%CITATION = doi:10.1103/PhysRevD.94.083519;%%
  %7 citations counted in INSPIRE as of 08 Mar 2017
  
  \bibitem{Zhang:2015uhk}
  X.~Zhang,
  ``Impacts of dark energy on weighing neutrinos after Planck 2015,''
  Phys.\ Rev.\ D {\bf 93}  n.~8,  083011 (2016)
%  doi:10.1103/PhysRevD.93.083011
  [arXiv:1511.02651 [astro-ph.CO]].
  %%CITATION = doi:10.1103/PhysRevD.93.083011;%%
  %16 citations counted in INSPIRE as of 08 Mar 2017

% mnu da HI
%\cite{Villaescusa-Navarro:2015cca}
%\bibitem{Villaescusa-Navarro:2015cca} 
%  F.~Villaescusa-Navarro, P.~Bull and M.~Viel,
%  ``Weighing neutrinos with cosmic neutral hydrogen,''
%  Astrophys.\ J.\  {\bf 814}, no. 2, 146 (2015)
%  doi:10.1088/0004-637X/814/2/146
%  [arXiv:1507.05102 [astro-ph.CO]].
  %%CITATION = doi:10.1088/0004-637X/814/2/146;%%
  %15 citations counted in INSPIRE as of 10 Feb 2017

% simulations kSZ
%\cite{Roncarelli:2017cwe}
%\bibitem{Roncarelli:2017cwe} 
%  M.~Roncarelli, F.~Villaescusa-Navarro and M.~Baldi,
%  ``The kinematic Sunyaev-Zel'dovich effect of the large-scale structure (I): dependence on neutrino mass,''
%  doi:10.1093/mnras/stx170
%  arXiv:1702.00676 [astro-ph.CO].
  %%CITATION = doi:10.1093/mnras/stx170;%%
    
%\cite{Aghanim:2015xee}
\bibitem{Aghanim:2015xee} 
  N.~Aghanim {\it et al.} [Planck Collaboration],
  ``Planck 2015 results. XI. CMB power spectra, likelihoods, and robustness of parameters,''
  %Submitted to: Astron. Astrophys.
  [arXiv:1507.02704 [astro-ph.CO]].
 
    \bibitem{newtau}
% \bibitem{Aghanim:2016yuo}
  N.~Aghanim {\it et al.} [Planck Collaboration],
  ``Planck 2016 intermediate results. XLVI. Reduction of large-scale systematic effects in HFI polarization maps and estimation of the reionization optical depth,''
  arXiv:1605.02985 [astro-ph.CO].
  %%CITATION = ARXIV:1605.02985;%%
  %6 citations counted in INSPIRE as of 24 May 2016

\bibitem{beutler2011}
  F.~Beutler
%, C.~Blake, M.~Colless, D.~H.~Jones, L.~Staveley-Smith, L.~Campbell, Q.~Parker and W.~Saunders 
{\it et al.},
  ``The 6dF Galaxy Survey: Baryon Acoustic Oscillations and the Local Hubble Constant,''
  Mon.\ Not.\ Roy.\ Astron.\ Soc.\  {\bf 416}, 3017 (2011)
  [arXiv:1106.3366 [astro-ph.CO]].
  %%CITATION = ARXIV:1106.3366;%%
  %406 citations counted in INSPIRE as of 04 Jun 2015

\bibitem{ross2014}
  A.~J.~Ross
%, L.~Samushia, C.~Howlett, W.~J.~Percival, A.~Burden and M.~Manera,
{\it et al.},
  ``The Clustering of the SDSS DR7 Main Galaxy Sample I: A 4 per cent Distance Measure at z=0.15,''
  Mon.\ Not.\ Roy.\ Astron.\ Soc.\  {\bf 449}, 835 (2015)
  [arXiv:1409.3242 [astro-ph.CO]].
  %%CITATION = ARXIV:1409.3242;%%
  %17 citations counted in INSPIRE as of 04 juin 2015

\bibitem{anderson2014}
  L.~Anderson {\it et al.}  [BOSS Collaboration],
  ``The clustering of galaxies in the SDSS-III Baryon Oscillation Spectroscopic Survey: baryon acoustic oscillations in the Data Releases 10 and 11 Galaxy samples,''
  Mon.\ Not.\ Roy.\ Astron.\ Soc.\  {\bf 441} n.~1,  24 (2014)
  [arXiv:1312.4877 [astro-ph.CO]].
  %%CITATION = ARXIV:1312.4877;%%


%\cite{Ade:2015zua}
\bibitem{Ade:2015zua} 
  P.~A.~R.~Ade {\it et al.} [Planck Collaboration],
  ``Planck 2015 results. XV. Gravitational lensing,''
    Astron.\ Astrophys.\  {\bf 594}, A15 (2016)
%  doi:10.1051/0004-6361/201525941
  [arXiv:1502.01591 [astro-ph.CO]].
  %%CITATION = doi:10.1051/0004-6361/201525941;%%
  %116 citations counted in INSPIRE as of 18 Jul 2016

\bibitem{cala}
  E.~Calabrese, A.~Slosar, A.~Melchiorri, G.~F.~Smoot and O.~Zahn,
  ``Cosmic Microwave Weak lensing data as a test for the dark universe,''
  Phys.\ Rev.\ D {\bf 77}, 123531 (2008)
  [arXiv:0803.2309 [astro-ph]].
  %%CITATION = ARXIV:0803.2309;%%
  %35 citations counted in INSPIRE as of 04 Jun 2015

  %\cite{Lewis:2002ah}
\bibitem{Lewis:2002ah} 
  A.~Lewis and S.~Bridle,
  ``Cosmological parameters from CMB and other data: A Monte Carlo approach,''
  Phys.\ Rev.\ D {\bf 66}, 103511 (2002)
  [astro-ph/0205436].

%\cite{Lewis:1999bs}
\bibitem{Lewis:1999bs} 
  A.~Lewis, A.~Challinor and A.~Lasenby,
  ``Efficient computation of CMB anisotropies in closed FRW models,''
  Astrophys.\ J.\  {\bf 538}, 473 (2000)
  %doi:10.1086/309179
  [astro-ph/9911177].
  %%CITATION = doi:10.1086/309179;%%
  %2057 citations counted in INSPIRE as of 19 Oct 2016
  
%\cite{Lewis:2013hha}
\bibitem{Lewis:2013hha} 
  A.~Lewis,
  ``Efficient sampling of fast and slow cosmological parameters,''
  Phys.\ Rev.\ D {\bf 87}, no. 10, 103529 (2013)
  [arXiv:1304.4473 [astro-ph.CO]].

%%%%%%%%%%%%%%%%%%%%%%%%%%%%%%%%%%%%%%%%%%%%%%%%%%%%%%%%%%%%%%%%%%%%%%%%%%
%%%%%%%%%%%%%%%%%%%%%%%%%%%%%%%%%%%%%%%%%%%%%%%%%%%%%%%%%%%%%%%%%%%%%%%%
    
    
\bibitem{Neym67}    
J.\ Neyman, ``Outline of a Theory of Statistical Estimation Based on the Classical Theory of Probability,''
Phil.\ Trans.\ Royal Soc.\ London, Series {\bf A 236}, 333 (1937), reprinted in ``A Selection of Early Statistical Papers on J. Neyman,'' (University of California Press, Berkeley, 1967).    
    
        
\bibitem{Feld98} 
  G.~J.~Feldman and R.~D.~Cousins,
  ``A Unified approach to the classical statistical analysis of small signals,''
  Phys.\ Rev.\ D {\bf 57}, 3873 (1998)
%  doi:10.1103/PhysRevD.57.3873
  [physics/9711021 [physics.data-an]].


\bibitem{Rodejo1}   
  W.~Rodejohann,
  ``Neutrinoless double beta decay and neutrino physics,''
  J.\ Phys.\ G {\bf 39}, 124008 (2012)
%  doi:10.1088/0954-3899/39/12/124008
  [arXiv:1206.2560 [hep-ph]].   


\bibitem{DellOro1}
  S.~Dell'Oro, S.~Marcocci and F.~Vissani,
  ``New expectations and uncertainties on neutrinoless double beta decay,''
  Phys.\ Rev.\ D {\bf 90}, no. 3, 033005 (2014)
%  doi:10.1103/PhysRevD.90.033005
  [arXiv:1404.2616 [hep-ph]].


\bibitem{DellOro2}
  S.~Dell'Oro, S.~Marcocci, M.~Viel and F.~Vissani,
  ``The contribution of light Majorana neutrinos to neutrinoless double beta decay and cosmology,''
  JCAP {\bf 1512}, no. 12, 023 (2015)
%  doi:10.1088/1475-7516/2015/12/023
  [arXiv:1505.02722 [hep-ph]].
  %%CITATION = doi:10.1088/1475-7516/2015/12/023;%%
    

\bibitem{Zavanin}     
  C.~Giunti and E.~M.~Zavanin,
  ``Predictions for Neutrinoless Double-Beta Decay in the 3+1 Sterile Neutrino Scenario,''
  JHEP {\bf 1507}, 171 (2015)
%  doi:10.1007/JHEP07(2015)171
  [arXiv:1505.00978 [hep-ph]].
    
%\cite{Xing:2016ymd}
\bibitem{Xing} 
  Z.~z.~Xing and Z.~h.~Zhao,
  ``The effective neutrino mass of neutrinoless double-beta decays: how possible to fall into a well,''
  arXiv:1612.08538 [hep-ph].
  %%CITATION = ARXIV:1612.08538;%%

\bibitem{Quiroga}
  H.~Minakata, H.~Nunokawa and A.~A.~Quiroga,
  ``Constraining Majorana CP phase in the precision era of cosmology and the double beta decay experiment,''
  PTEP {\bf 2015}, 033B03 (2015)
%  doi:10.1093/ptep/ptv010
  [arXiv:1402.6014 [hep-ph]].
  %%CITATION = doi:10.1093/ptep/ptv010;%%
 
\bibitem{Riotto} 
  S.~Pascoli, S.~T.~Petcov and A.~Riotto,
  ``Leptogenesis and Low Energy CP Violation in Neutrino Physics,''
  Nucl.\ Phys.\ B {\bf 774}, 1 (2007)
%  doi:10.1016/j.nuclphysb.2007.02.019
  [hep-ph/0611338].
  %%CITATION = doi:10.1016/j.nuclphysb.2007.02.019;%%
   
    
\bibitem{Katrin1} J.\ Angrik et al.\ [KATRIN Collaboration], KATRIN Design Report 2004 (245 pages), 
 Report FZKA-7090, NPI ASCR Rez EXP-01/2005, MS-KP-0501. Available at the website:
 www.katrin.kit.edu

\bibitem{Katrin2} 
  E.~W.~Otten and C.~Weinheimer,
  ``Neutrino mass limit from tritium beta decay,''
  Rept.\ Prog.\ Phys.\  {\bf 71}, 086201 (2008)
%  doi:10.1088/0034-4885/71/8/086201
  [arXiv:0909.2104 [hep-ex]].

\bibitem{Katrin3} G.\ Drexlin, talk at {\em NOW 2016\/} \protect\cite{SKatm2}. 

\bibitem{Gastaldo}  L.\ Gastaldo, talk at {\em NOW 2016\/} \protect\cite{SKatm2}, PoS(NOW2016)060.
    
\bibitem{Project8}     
  A.~A.~Esfahani {\it et al.},
  ``Determining the neutrino mass with Cyclotron Radiation Emission Spectroscopy - Project 8,''
%  doi:10.1088/1361-6471/aa5b4f
  arXiv:1703.02037 [physics.ins-det].

    
    
    
    
    
\end{thebibliography}
\end{document}